# Detection of a Pair Density Wave State in UTe$_2$

Qiangqiang Gu[1§], Joseph P. Carroll[1,2§], Shuqiu Wang[1,3§],
Sheng Ran[4], Christopher Broyles[4], Hasan Siddiquee[4],
Nicholas P. Butch[5,6], Shanta R. Saha[5], Johnpierre Paglione[5,7],
J.C. Séamus Davis[1,2,3,8] and Xiaolong Liu[1,9,10]

1. LASSP, Department of Physics, Cornell University, Ithaca, NY 14850, USA
2. Department of Physics, University College Cork, Cork T12 R5C, IE
3. Clarendon Laboratory, University of Oxford, Oxford, OX1 3PU, UK
4. Department of Physics, Washington University. St Louis, MO 63130, USA
5. Maryland Quantum Materials Center, University of Maryland, College Park, MD 20742, USA
6. NIST Center for Neutron Research, 100 Bureau Drive, Gaithersburg MD 20899, USA
7. Canadian Institute for Advanced Research, Toronto, Ontario M5G 1Z8, Canada
8. Max-Planck Institute for Chemical Physics of Solids, D-01187 Dresden, DE
9. Department of Physics and Astronomy, University of Notre Dame, Notre Dame, IN 46556, USA
10. Stavropoulos Center for Quantum Matter, University of Notre Dame, Notre Dame, IN 46556, USA
§ These authors contributed equally to this project.

**Spin-triplet topological superconductors should exhibit many unprecedented electronic properties including fractionalized electronic states relevant to quantum information processing. Although UTe$_2$ may embody such bulk topological superconductivity[1-11], its superconductive order-parameter $\Delta(k)$ remains unknown[12]. Many diverse forms for $\Delta(k)$ are physically possible[12] in such heavy fermion materials[13]. Moreover, intertwined[14,15] density waves of spin (SDW), charge (CDW) and pairs (PDW) may interpose, with the latter exhibiting spatially modulating[14,15] superconductive order-parameter $\Delta(r)$, electron pair density[16-19] and pairing energy-gap[17,20-23]. Hence, the newly discovered CDW state[24] in UTe$_2$ motivates the prospect that a PDW state may exist in this material[24,25]. To search for it, we visualize the pairing energy-gap with µeV − scale energy-resolution using superconductive STM tips[26-31]. We detect three PDWs, each with peak-peak gap modulations circa 10 µeV and at incommensurate wavevectors $P_{i=1,2,3}$ that are indistinguishable from the wavevectors $Q_{i=1,2,3}$ of the prevenient[24] CDW. Concurrent visualization of the UTe$_2$ superconductive PDWs and the non-superconductive CDWs reveals that every $P_i$: $Q_i$ pair exhibits a relative spatial phase $\delta\phi \approx \pi$. From these observations and given UTe$_2$ as a spin-triplet superconductor[12], this PDW state should be a spin-triplet pair density wave[24,25]. While such states do exist[32] in superfluid $^3$He, for superconductors they are unprecedented.**

**Spin-triplet pair density waves**

Bulk Cooper-pair condensates are definitely topological when their superconductive or superfluid order-parameters exhibit odd parity [33,34] $\Delta(\boldsymbol{k}) = -\Delta(-\boldsymbol{k})$ with spin-triplet pairing. This situation is epitomized by liquid $^3$He, the only known bulk topological Cooper-pair condensate [35,36]. Although no bulk superconductor exhibits an unambiguously topological $\Delta(\boldsymbol{k})$, attention has recently fixed upon the compound UTe$_2$ as a promising candidate[1-12]. This material is superconducting below the critical temperature $T_c$ = 1.65 K. Its extremely high critical magnetic field and the minimal suppression of the Knight shift[3] upon entering the superconductive state, both imply spin-triplet superconductivity[1,2]. Temperature[4], magnetic field[4,5] and angular dependence[5] of the superconductive quasiparticle thermal conductivity are all indicative of a superconducting energy gap with point nodes[4-6]. In the superconductive phase, evidence for time-reversal symmetry breaking is provided by polar Kerr rotation measurements[7] but is absent in muon spin rotation studies[8]. Additionally, the superconductive electronic structure when visualized at opposite mesa-edges at the UTe$_2$ (0-11) surface, breaks chiral symmetry[9]. Dynamically, UTe$_2$ appears to contain both strong ferromagnetic and antiferromagnetic spin fluctuations[10,11] relevant to superconductivity. Together these results are consistent with a spin-triplet and thus odd-parity, nodal, time-reversal symmetry breaking, chiral superconductor[12]. Figure 1a shows a schematic of the crystal structure of this material, while Fig. 1c is a schematic of the Fermi surface in the ($k_x$, $k_y$) plane at $k_z = 0$ (dashed lines, Ref. 37). An exemplary order-parameter $\Delta(\boldsymbol{k})$ hypothesized[5] for UTe$_2$ is also shown schematically in Fig. 1c (solid lines); but numerous others have been proposed[12], including that of a PDW state[24,25]. In theory, this PDW if generated by time-reversal and surface reflection symmetry breaking, is a spin-triplet pair density wave[25]. Such a state is unknown for superconductors but occurs in topological superfluid $^3$He[32].

**Pair density wave visualization**

In general, a PDW state is a superconductor but with a spatially modulating superconductive order-parameter[14,15]. Absent flowing currents or magnetic fields, a conventional spin-singlet superconductor has an order-parameter

*2*

$$\Delta_S(\boldsymbol{r}) = \Delta_0 e^{i\phi_S} \qquad (1)$$

for which $\phi_S$ is the macroscopic quantum phase and $\Delta_0$ the amplitude of the many-body condensate wavefunction. A unidirectional PDW modulates such an order-parameter at wavevector $\boldsymbol{P}$ as

$$\Delta_P(\boldsymbol{r}) = \Delta(\boldsymbol{r})e^{i\boldsymbol{P}\cdot\boldsymbol{r}} + \Delta^*(\boldsymbol{r})e^{-i\boldsymbol{P}\cdot\boldsymbol{r}} \qquad (2)$$

meaning that the electron-pairing potential undulates spatially. By contrast, a unidirectional CDW modulates the charge density at wavevector $\boldsymbol{Q}$ such that

$$\rho_Q(\boldsymbol{r}) = \rho(\boldsymbol{r})e^{i\boldsymbol{Q}\cdot\boldsymbol{r}} + \rho^*(\boldsymbol{r})e^{-i\boldsymbol{Q}\cdot\boldsymbol{r}} \qquad (3)$$

The simplest interactions between these three orders can be analyzed using a Ginzburg-Landau-Wilson (GLW) free energy density functional

$$\mathcal{F} = \lambda[\rho_Q \Delta_S^* \Delta_P + c.c.] \qquad (4)$$

representing the lowest order coupling between superconductive and density wave states. There are two elementary possibilities: (a) if $\Delta_S(\boldsymbol{r})$ and $\Delta_P(\boldsymbol{r})$ are the predominant orders, they generate charge modulations of form $\rho_P(\boldsymbol{r}) \propto \Delta_S^* \Delta_P + \Delta_{-P}^* \Delta_S$ and $\rho_{2P}(\boldsymbol{r}) \propto \Delta_{-P}^* \Delta_P$, i.e. two induced CDWs controlled by the wavevector of the PDW; (b) if $\Delta_S(\boldsymbol{r})$ and $\rho_Q(\boldsymbol{r})$ are predominant orders, they generate modulations $\Delta_Q(\boldsymbol{r}) \propto \Delta_S^* \rho_Q$, i.e. a PDW induced at the wavevector of the CDW. In either case, the PDW state described by Eqn. 2 subsists.

To explore UTe$_2$ for such physics, it is first necessary to simultaneously visualize any coexisting CDW and PDW states. Recent experimental advances have demonstrated two techniques for visualizing a PDW state. In the first[16-19], the condensed electron-pair density at location $\boldsymbol{r}$, $n(\boldsymbol{r})$, can be visualized by measuring the tip-sample Josephson critical-current squared $I_J^2(\boldsymbol{r})$ from which

$$n(\boldsymbol{r}) \propto I_J^2(\boldsymbol{r}) R_N^2(\boldsymbol{r}) \qquad (5)$$

where $R_N(\boldsymbol{r})$ is the normal-state junction resistance. In the second PDW visualization technique[17,20-23], the magnitude of the total energy gap in the sample $\Delta(\boldsymbol{r})$ is defined as half the energy separation between the two superconductive coherence peaks in the density of electronic states $N(E)$. These occur in tunneling conductance at energies $\Delta_+(\boldsymbol{r})$ and $\Delta_-(\boldsymbol{r})$ such that

$$\Delta(\boldsymbol{r}) \equiv [\Delta_+(\boldsymbol{r}) - \Delta_-(\boldsymbol{r})]/2 \tag{6}$$

This can be visualized using either normal-insulator-superconductor (NIS) tunneling[20-22], or superconductor-insulator-superconductor (SIS) tunneling from a superconductive scanning tunneling microscopy (STM) tip[17,23,29] whose superconductive gap energy $\Delta_{\text{tip}}$ is known *a priori*.

**CDW visualization in normal-state UTe$_2$**

UTe$_2$ crystals typically cleave to reveal the (0-11) surface[9,24], a schematic view of which (Fig. 1b) identifies the key atomic periodicities by vectors ***a*****, *b****. At temperature $T$ = 4.2 K, this surface is visualized using STM and a typical topographic image $T(\boldsymbol{r})$ is shown in Fig. 1d, while Fig. 1e shows its power spectral density Fourier transform $T(\boldsymbol{q})$ with the surface reciprocal-lattice points identified by orange dashed circles. Pioneering STM studies of UTe$_2$ by Aishwarya *et al.*[24] have recently discovered a CDW state by visualizing the electronic density-of-states $g(\boldsymbol{r}, E)$ of such surfaces. As well as the standard maxima at the surface reciprocal-lattice points in $g(\boldsymbol{q}, E)$, the Fourier transform of $g(\boldsymbol{r}, E)$, Aishwarya *et al.* detected three new maxima with incommensurate wavevectors $\boldsymbol{Q}_{1,2,3}$ signifying the existence of a CDW state occurring at temperatures up to at least $T$ = 10 K. To emulate, we measure $g(\boldsymbol{r}, V)$ for -25 mV $< V <$ 25 mV at $T$ = 4.2 K using a non-superconducting tip on the equivalent cleave surface to Ref. 24. Figure 2a shows a typical topographic image $T(\boldsymbol{r})$ of the (0-11) surface measured at 4.2 K. The Fourier transform $T(\boldsymbol{q})$ features the surface reciprocal-lattice points labeled by orange dashed circles in Fig. 2a inset. The simultaneous image $g(\boldsymbol{r}, 10\text{ mV})$ in Fig. 2b exhibits the typical modulations in $g(\boldsymbol{r}, V)$ and its Fourier transform $g(\boldsymbol{q}, V)$ in Fig. 2c reveals the three CDW peaks[24] at $\boldsymbol{Q}_{1,2,3}$ labeled by blue dashed circles. Inverse Fourier filtration of these three maxima only, reveals the incommensurate CDW state of UTe$_2$. Overall, this state consists predominantly of incommensurate charge density modulations at three (0-11) in-plane wavevectors $\boldsymbol{Q}_{1,2,3}$ that occur at temperatures up to at least 10 K[24], and with a characteristic energy scale up to at least $\pm$25 meV (Ref. 24, Methods and Extended Data Fig. 1).

**Normal-tip PDW detection at NIS gap-edge**

Motivated by the discovery that this CDW exhibits an unusual dependence on magnetic field and by the consequent hypothesis that a PDW may exist in this material[24,25], we next consider direct PDW detection in UTe2 by visualizing spatial modulations in its energy gap[17,18,20-23]. The typical tunneling conductance signature of the UTe2 superconducting energy-gap is exemplified in Fig. 3a, showing a density-of-states spectrum $N(E = \text{eV}) \propto dI/dV|_{\text{NIS}}(V)$ measured using a non-superconducting tip at $T$ = 280 mK and junction resistance of $R \approx 5$ MΩ. Under these circumstances, researchers find only a small drop in the tunneling conductance at energies $|E| \leq |\Delta_{\text{UTe}_2}|$ [9] and concomitantly weak energy-maxima in $N(E)$ at the energy gap edges $E \approx \pm\Delta_{\text{UTe}_2}$ (Fig. 3a inset). Hence, it is challenging to determine accurately the precise value of the energy gap $\Delta_{\text{UTe}_2}$ (Methods and Extended Data Fig. 3). Nevertheless, we fit a second-order polynomial to the two energy-maxima in measured $N(E, \boldsymbol{r})$ surrounding $E \approx \pm\Delta_{\text{UTe}_2}$, evaluate the images $\Delta_\pm(\boldsymbol{r})$ of these energies and then derive a gap map for UTe2 as $\Delta_{\text{UTe}_2}(\boldsymbol{r}) \equiv [\Delta_+(\boldsymbol{r}) - \Delta_-(\boldsymbol{r})]/2$. Its Fourier transform $\Delta_{\text{UTe}_2}(\boldsymbol{q})$ presented in Methods and Extended Data Fig. 3 reveals three incommensurate energy gap modulations occurring at wavevectors $\boldsymbol{P}_{i=1,2,3}$ consistent with the wavevectors of the CDW modulations discovered in Ref. 24. While this evidence of three PDW states in UTe2 is encouraging, its weak signal-to-noise-ratio due to the shallowness of coherence peaks implies that conventional $dI/dV|_{\text{NIS}}$ spectra are inadequate for precision application of Eqn. 6 in this material.

**Superconductive-tip PDW detection**

Therefore, we turn to a well-known technique for improving the resolution of energy-maxima in $N(E, \boldsymbol{r})$. By using SIS tunneling from a tip exhibiting high sharp conductance peaks, one can profoundly enhance energy resolution for quasiparticles[26-31]. Most recently this has been demonstrated in electronic fluid flow visualization[29] microscopy, with effective energy resolution $\delta E \approx 10$ μeV. The SIS current $I$ from a superconducting tip is given by the convolution

$$I(V) \propto \int_o^{eV} N_{\text{tip}}(E)\, N_{\text{sample}}(E - eV)\, dE \tag{7}$$

Equation 7 demonstrates that using a superconductive tip with high sharp coherence peaks at $E_\pm = \pm\Delta_{\text{tip}}$ in $N_{\text{tip}}(E)$ will, through convolution, strongly enhance the resolution for measuring the energies $\pm\Delta_{\text{sample}}$ at which energy-maxima occur in $N_{\text{sample}}(E)$; it will also shift the energy of these features to $E = \pm[\Delta_{\text{sample}} + \Delta_{\text{tip}}]$. In Fig. 3b we show the $dI/dV|_{\text{SIS}}$ spectrum of a UTe₂ single crystal using a superconducting Nb tip at $T$ = 280 mK. Because the tunneling current is given by Eqn. 7, the clear maxima in $dI/dV|_{\text{SIS}}$ occur at energies $\pm(\Delta_{\text{tip}} + \Delta_{\text{sample}})$. With this technique the energy-maxima can be identified with resolution better than $\delta E \leq 10$ μeV when $T$ < 300 mK[29]. Here we use it to improve the signal-to-noise ratio of the UTe₂ superconductive energy-gap modulations that are already detectable by conventional techniques (Methods and Extended Data Fig. 3).

The UTe₂ samples are cooled to $T$ = 280 mK, with $T(r, V)$ of the (0-11) cleave surface as measured by a superconductive Nb tip shown in Fig. 3c. Here we see a powerful enhancement in the amplitude and sharpness of maxima in $dI/dV|_{\text{SIS}}$ relative to Fig. 3a. Consequently, to determine the spatial arrangements of the energy of the two maxima $E_+(r), E_-(r)$ surrounding 1.6 meV exemplified in Fig. 3b, we make two separate $g(r, V)$ maps in the sample bias voltage $V$ ranges -1.68 mV < $V$ < -1.48 mV and 1.5 mV < $V$ < 1.7 mV, and in the identical field of view (FOV). The sharp peak of each $dI/dV|_{\text{SIS}}$ is fit to a second-order polynomial $dI/dV|_{\text{SIS}} = aV^2 + bV + c$, achieving typical quality of fit $R^2 = 0.99 \pm 0.005$. The energy of maximum intensity in $E_+(r)$ or $E_-(r)$ is then identified analytically from the fit parameters (Methods and Extended Data Fig. 4). The fine line across Fig. 3c specifies the trajectory of an exemplary series of $dI/dV|_{\text{SIS}}$ spectra, while Fig. 3d presents the colormap $dI/dV|_{\text{SIS}}$ spectra for both positive and negative energy coherence peaks along this line. Periodic variations in the energies at which pairs of peaks occur are obvious, directly demonstrating that $E_+(r)$ and $E_-(r)$ are modulating periodically but in energetically opposite directions. Using this $g(r, V)$ measurement and fitting procedure (Methods and Extended Data Fig. 4) yields atomically resolved images of $E_+(r)$ and $E_-(r)$. The magnitude of both positive and negative superconductive energy gaps of UTe₂ is then $\Delta_\pm(r) \equiv |E_\pm(r)| - |\Delta_{\text{tip}}|$ where $|\Delta_{\text{tip}}|$ is constant. These two independently measured gap maps $\Delta_+(r), \Delta_-(r)$ are spatially registered to each other at every location with 27 pm precision so that the cross-correlation coefficient between them is $X \cong 0.92$, meaning that the superconducting energy gap modulations are entirely particle-hole symmetric (Figs. 3e, f; Methods and Extended Data Fig. 5).

From these and equivalent data, the UTe$_2$ superconducting energy gap structure $\Delta_{\text{UTe}_2}(\boldsymbol{r}) = (\Delta_+(\boldsymbol{r}) + \Delta_-(\boldsymbol{r}))/2$ can now be examined for its spatial variations $\delta\Delta(\boldsymbol{r})$ by using

$$\delta\Delta(\boldsymbol{r}) \equiv \Delta_{\text{UTe}_2}(\boldsymbol{r}) - \langle\Delta_{\text{UTe}_2}(\boldsymbol{r})\rangle \tag{8}$$

where $\langle\Delta_{\text{UTe}_2}(\boldsymbol{r})\rangle$ is the spatial average over the whole FOV. Figure 4a shows measured $\delta\Delta(\boldsymbol{r})$ in the identical FOV to Fig. 3c. The Fourier transform of $\delta\Delta(\boldsymbol{r})$, $\delta\Delta(\boldsymbol{q})$, is presented in Fig. 4b in which the surface reciprocal-lattice points are identified by orange dashed circles. The three additional peaks labeled by red dashed circles represent energy gap modulations with incommensurate wavevectors at $\boldsymbol{P}_{1,2,3}$ of the PDW state in UTe$_2$. Focusing only on these three wavevectors $\boldsymbol{P}_{1,2,3}$ we perform an inverse Fourier transform to reveal the spatial structure of the UTe$_2$ PDW state in Fig. 4c (Methods). This state appears to consist predominantly of incommensurate superconductive energy gap modulations at three (0-11) in-plane wavevectors $\boldsymbol{P}_{1,2,3}$ with a characteristic energy scale 10 μeV for peak-peak modulations.

**Energy modulations of Andreev resonances**

There is an alternative modality of SIS tunneling, namely measuring the effects of multiple Andreev reflections. For two superconductors with very different gap magnitudes, when the sample bias voltage shifts the smaller gap edge (UTe$_2$ in this case) to the chemical potential of the other superconductor, the Andreev process of electron (hole) transmission and hole (electron) reflection plus electron-pair propagation can produce an energy maximum in $dI/dV|_{\text{SIS}}$[38], an effect well attested by experiment[39]. Here, by imaging the energies of $A_\pm(\boldsymbol{r})$ of two subgap $dI/dV|_{\text{SIS}}$ maxima detected throughout our studies and identified by green arrows in Fig. 3b, an Andreev-resonance measure of the UTe$_2$ energy gap is conjectured as $\Delta_A(\boldsymbol{r}) \equiv [A_+(\boldsymbol{r})-A_-(\boldsymbol{r})]/2$. These data are presented in Methods and Extended Data Fig. 7 and reveal a $\Delta_A(\boldsymbol{r})$ modulating with amplitude approximately 10 μeV at wavevectors $\boldsymbol{P}_{1,2}$ state, further evidencing the UTe$_2$ PDW state.

**Visualizing interplay of PDW and CDW**

Finally, one may consider the two cases of intertwining outlined earlier: (a) $\Delta_S(\boldsymbol{r})$ and $\Delta_P(\boldsymbol{r})$ are predominant and generate charge modulations $\rho_P(\boldsymbol{r}) \propto \Delta_S^* \Delta_P + \Delta_{-P}^* \Delta_S$ and $\rho_{2P}(\boldsymbol{r}) \propto \Delta_{-P}^* \Delta_P$ or, (b) $\Delta_S(\boldsymbol{r})$ and $\rho_Q(\boldsymbol{r})$ are predominant and generate pair density modulations $\Delta_Q(\boldsymbol{r}) \propto \Delta_S^* \rho_Q$. For case (a) to be correct, a PDW with magnitude 10 μeV coexisting with a superconductor of gap maximum near 250 μeV must generate a CDW on the energy scale 25 meV and exist up to at least $T$ = 10 K. For case (b) to be valid, a normal-state CDW with eigenstates at energies up to 25 meV coexisting with a superconductor of gap magnitude 250 μeV must generate a PDW at the same wavevector and with amplitude near 10 μeV. Intuitively, the latter case seems the most plausible for UTe$_2$.

To explore this issue further, we visualize the CDW in the non-superconductive state at $T$ = 4.2 K, then cool to $T$ = 280 mK and visualize the PDW in precisely the same FOV. Figures 4c, d show the result of such an experiment in the FOV of Fig. 3c. The CDW and PDW images are registered to the underlying lattice and to each other with 27 pm precision. Comparing their coterminous images in Fig. 4c and Fig. 4d reveals that the CDW and PDW states of UTe$_2$ appear spatially quite distinct. Yet, they are actually registered to each other in space, being approximate negative images of each other (Fig. 4e) with a measured relative phase for all three $\boldsymbol{P}_i:\boldsymbol{Q}_i$ pairs of $|\delta\phi_i| \cong \pi$ (Fig. 4f, Methods and Extended Data Fig. 10). A typical example of this effect is revealed in a linecut across Figs. 4c, d along the Te chain direction, with the directly measured values shown in Fig. 4g. The direct and comprehensive knowledge of CDW and PDW characteristics and interactions presented in Fig. 4, now motivates search for a Ginzburg-Landau description capable of capturing this complex intertwined phenomenology and that reported in Ref. 24.

**Conclusions**

Notwithstanding such theoretical challenges, in this study we have demonstrated that PDWs occur at three incommensurate wavevectors $\boldsymbol{P}_{i=1,2,3}$ in UTe$_2$ (Figs. 4b, c). These wavevectors are indistinguishable from the wavevectors $\boldsymbol{Q}_{i=1,2,3}$ of the prevenient normal-state CDW (Figs. 2c, 4d). All three PDWs exhibit peak-peak gap energy modulations in the range near 10 µeV (Figs. 4c, g). When the $\boldsymbol{P}_{i=1,2,3}$ PDW states are visualized at 280 mK in the identical FOV as the $\boldsymbol{Q}_{i=1,2,3}$ CDWs visualized above the superconductive $T_c$, every $\boldsymbol{Q}_i:\boldsymbol{P}_i$ pair is spatially registered to each other (Figs. 4c,d), but with a relative phase shift of $|\delta\phi_i| \cong \pi$ throughout (Fig. 4f). Given the premise that UTe$_2$ is a spin-triplet superconductor[12], the PDW phenomenology detected and described herein (Fig. 4) signifies the entrée to spin-triplet pair density wave physics.

Figure 1

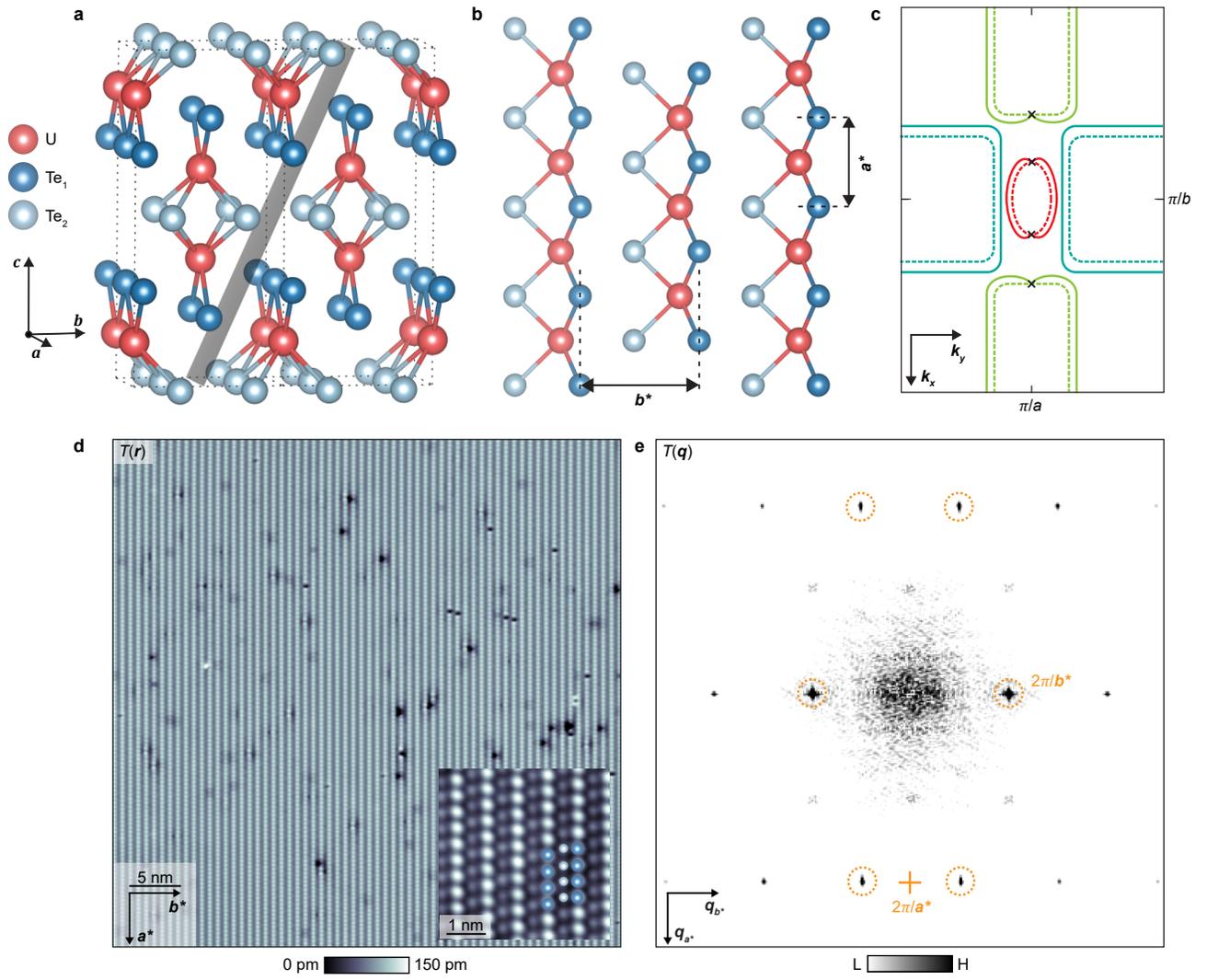

Figure 2

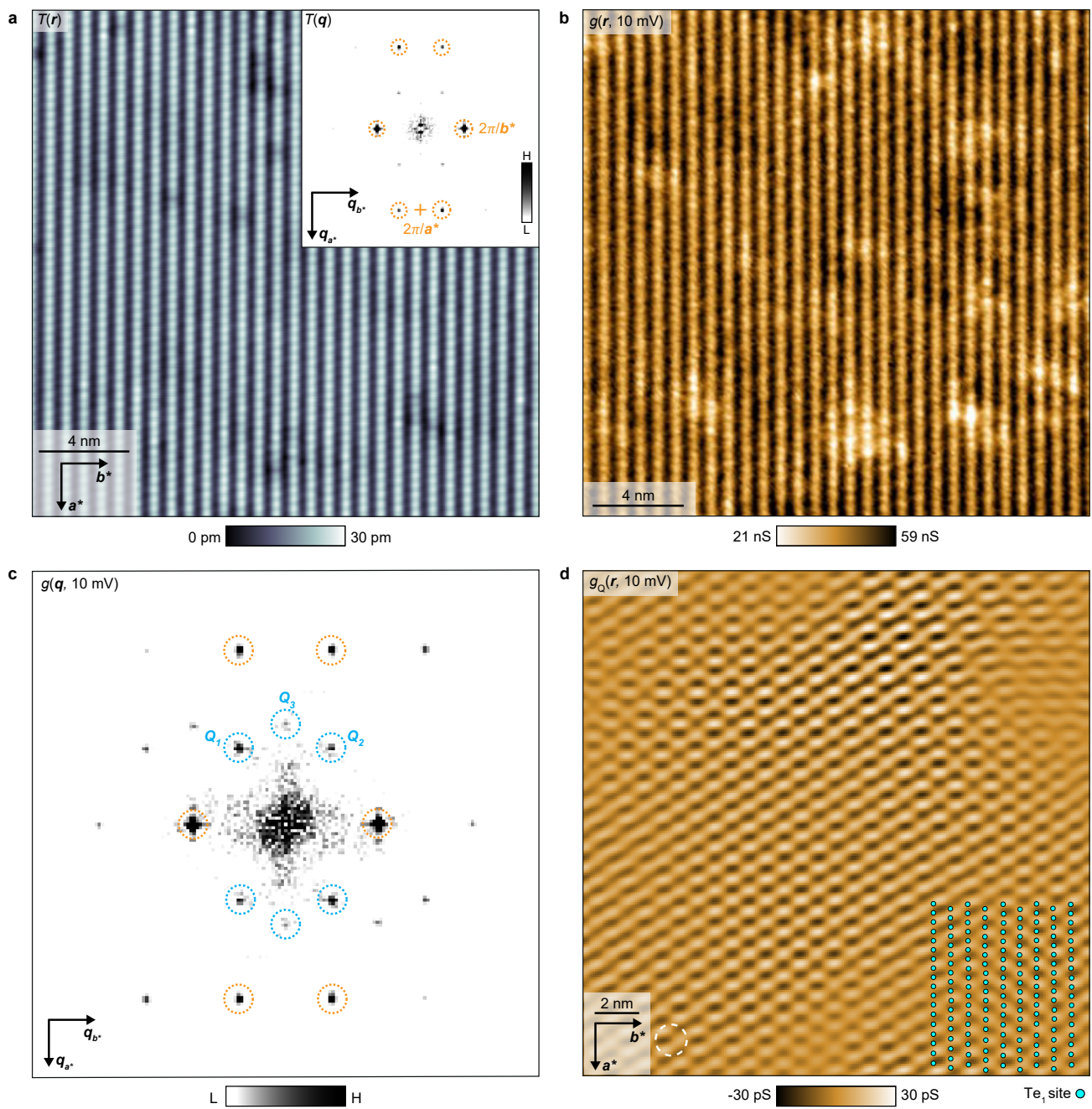

Figure 3

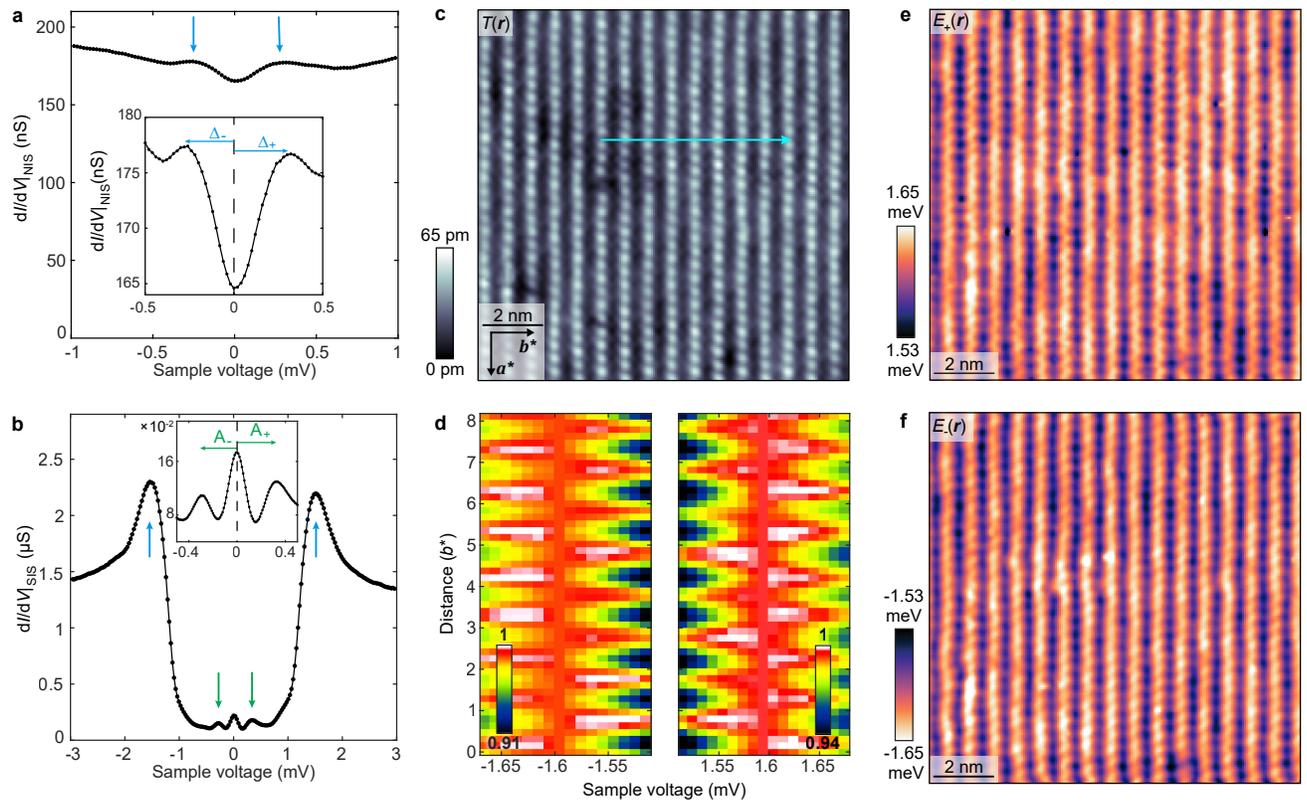

Figure 4

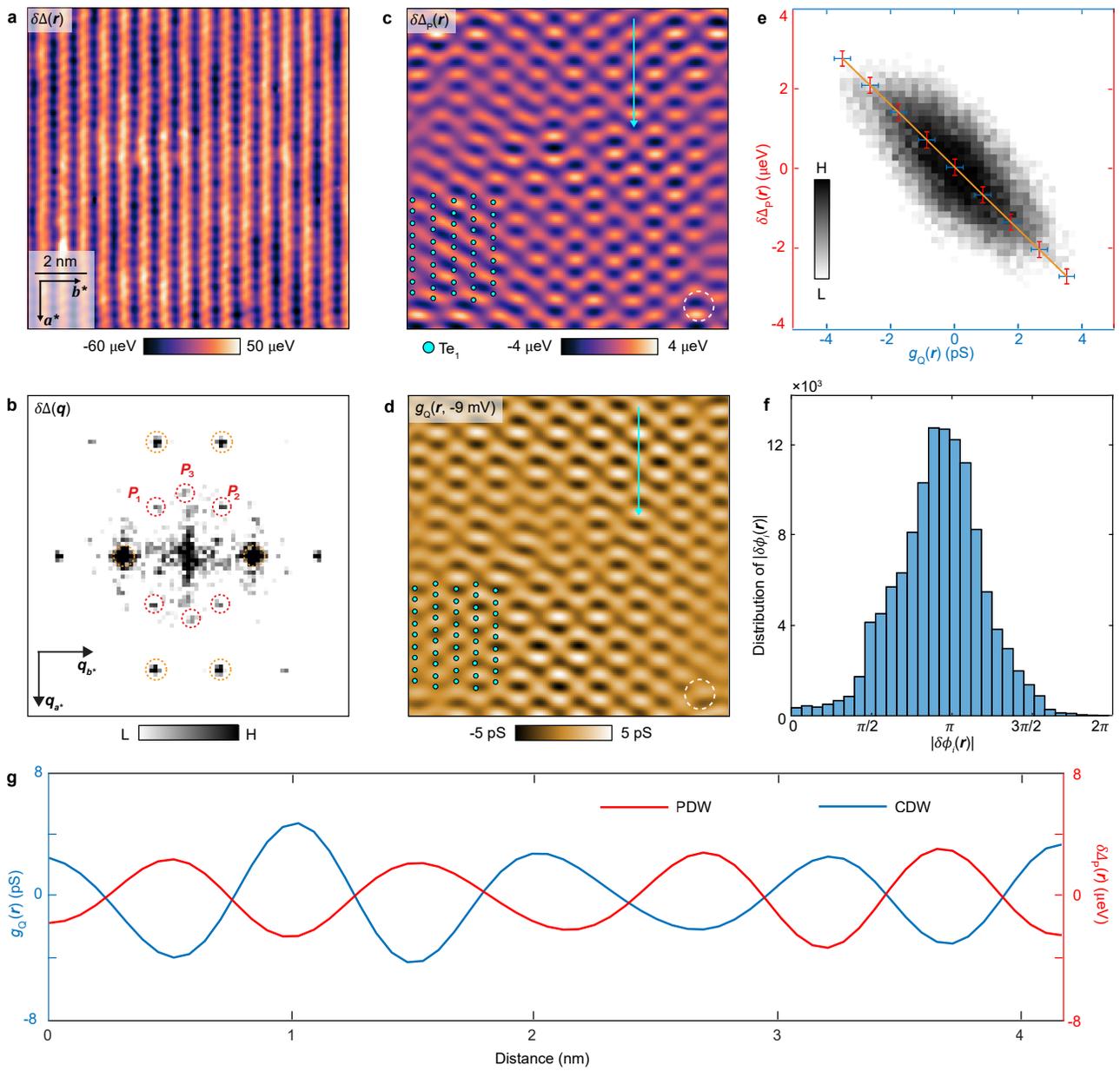

# Main Figure Legends

**FIG. 1 Momentum-Space and Real-Space Characteristics of UTe$_2$**

a. Schematic crystal lattice structure of UTe$_2$ oriented to the primary unit cell vectors **a**, **b**, **c**. The (0-11) cleave plane of UTe$_2$ is indicated schematically by the grey shaded plane.

b. Schematic of elemental identities and atomic sites and unit-cell of the (0-11) termination layer of cleaved UTe$_2$.

c. Schematic Fermi surface in the ($k_x$, $k_y$) plane at $k_z = 0$ for UTe$_2$ is indicated by dashed curves. A schematic example of one possible superconductive order parameter magnitude is indicated by solid curves representing the magnitude of energy gap $\Delta(\mathbf{k})$. Here for pedagogic purposes only, is presented a chiral, spin-triplet superconductor with energy gap nodes along the **a**-axis or $k_x$-axis.

d. Typical topographic image $T(\mathbf{r})$ of UTe$_2$ (0-11) surface measured with a superconducting tip at $T$ = 280 mK ($I_s$ = 0.5 nA, $V_s$ = 30 mV). Inset: measured high-resolution $T(\mathbf{r})$ at low junction resistance ($I_s$ = 3 nA, $V_s$ = 5 mV), clarifying two types of Te atoms.

e. Measured $T(\mathbf{q})$, the Fourier transform of $T(\mathbf{r})$ in **d**, with the surface reciprocal-lattice points labeled as dashed circles, which are consistent with simulated results (Extended Data Fig. 2).

**FIG. 2. Visualizing the Normal-State CDW of UTe$_2$**

a. Typical topographic image $T(\mathbf{r})$ of (0-11) surface measured at 4.2 K with a non-superconducting STM tip ($I_s$ = 1 nA, $V_s$ = -30 mV). Inset: measured $T(\mathbf{q})$, the Fourier transform of the topographic image obtained simultaneously as **b**. Reciprocal-lattice points labelled in orange dashed circles.

b. Differential conductance image $g(\mathbf{r}, 10\ \text{mV})$ measured at 4.2 K.

c. Fourier transform $g(\mathbf{q}, 10\ \text{mV})$ from $g(\mathbf{r}, 10\ \text{mV})$ in **b**. Three incommensurate CDW peaks at $\mathbf{Q}_{1,2,3}$ labeled by blue dashed circles.

d. Measured density-of-states modulations $g_Q(\mathbf{r}, 10\ \text{mV})$ only at the wavevectors $\mathbf{Q}_{1,2,3}$. This is a highly typical image of the incommensurate CDW state of UTe$_2$ (Methods and Extended Data Fig. 1). Te$_1$ atomic locations of UTe$_2$ (0-11) surface shown as overlay. The filter size of inverse Fourier transform is 14 Å.

**FIG. 3 Atomic-Resolution Imaging of the Superconductive Energy Gap**

a. Typical NIS spectrum $dI/dV|_{\text{NIS}}$ from normal-tip to UTe$_2$ (0-11) surface ($I_s$ = 1 nA, $V_s$ = -5 mV) at $T$ = 280 mK. The inset focuses on the energy range where the coherence peaks can be detected by conventional normal-tip tunneling at $E = \pm\Delta_{\text{UTe}_2}$. Visualizing the superconducting energy gap $\Delta_{\text{UTe}_2}(r)$ from such $dI/dV|_{\text{NIS}}$ imaging at $T$ = 280 mK reveals three sets of energy gap modulations occurring at PDW wavevectors $P_{i=1,2,3}$ (Methods and Extended Data Fig. 3). We find no deterministic influence of the residual density of states modulations on these PDW energy gap modulations (Methods and Extended Data Fig. 9).

b. Typical SIS spectrum $dI/dV|_{\text{SIS}}$ from superconducting Nb tip to UTe$_2$ (0-11) surface. The blue arrows indicate the convoluted conductance peak located at $|\Delta_{\text{tip}} + \Delta_{\text{UTe}_2}|$ ($I_s$ = 3 nA, $V_s$ = 3 mV). The inset focuses the energy range where subgap $dI/dV|_{\text{SIS}}(r,V)$ peaks can be detected at energies $E = A_\pm(r)$.

c. Typical SIS tunneling topograph $T(r)$ measured at $T$ = 280 mK.

d. Exemplary normalized $dI/dV|_{\text{SIS}}(r,V)$ focused on the energy ranges near $E_+$ and $E_-$ along the trajectory indicated as the light blue arrow in **c**. The modulations of the energies $E_+(r)$, $E_-(r)$ of maximum conductance are clearly seen. The two sets of spectra are calibrated such that the $dI/dV|_{\text{SIS}}(V)$ peaks are particle-hole symmetric.

e. Measured energy $E_+(r)$ at which $dI/dV|_{\text{SIS}}(V_+)$ maxima occur in **c**. The UTe$_2$ empty-state superconductive energy gap is $\Delta_+(r) = |E_+(r)| - |\Delta_{\text{tip}}|$ where $|\Delta_{\text{tip}}|$ is a constant.

f. Measured energy $E_-(r)$ at which $dI/dV|_{\text{SIS}}(V_-)$ maxima occur in **c**. The filled-state energy gap is $\Delta_-(r) = |E_-(r)| - |\Delta_{\text{tip}}|$.

**FIG. 4 Visualizing the Pair Density Wave State of UTe₂**

a. Measured variations in energy-gap $\delta\Delta(r)$ from 3**c**.

b. Measured $\delta\Delta(q)$ from **a**. The surface reciprocal-lattice points are labeled by orange circles and the PDW peaks at $P_{1,2,3}$ labeled in red. $P_{1,2,3}$ are linked by reciprocal-lattice vectors (Extended Data Fig. 10). $\delta\Delta(r)$ and $\delta\Delta(q)$ exhibit superior signal-to-noise ratio as compared to the normal-tip gap-map $\Delta_{\mathrm{UTe_2}}(r)$ (Extended Data Fig. 8).

c. Inverse Fourier transform (IFT) filtered $\delta\Delta(q)$ of **a** at $P_{1,2,3}$ reveals the first visualization of the PDW (filter size is 11.4 Å). The PDW is repeatable in experimental measurements (Extended Data Fig. 6) and also independently evidenced in Methods and Extended Data Fig. 7.

d. Image of $g_Q(r, -9\text{ mV})$ of the CDW, measured at $T$ = 4.2 K in the same FOV as **c** from IFT filtered $g(r, -9\text{ mV})$ at $Q_{1,2,3}$ (filter size is 11.4 Å). The precision of registration between the CDW and PDW images is 27 pm (Methods and Extended Data Fig. 5). These coincident CDW and PDW images are measured in the energy ranges $10\text{ meV}$ and $10\text{ μeV}$, respectively, and appear visually distinct, yet their cross-correlation coefficient $-0.68$ reveals their anticorrelation. The CDW maxima exist at the PDW minima.

e. Statistical relationship between $\delta\Delta_{\mathrm{P}}(r)$ and $g_Q(r, -9\text{ mV})$. The $\delta\Delta_{\mathrm{P}}(r)$ and $g_Q(r, -9\text{ mV})$ are strongly anticorrelated spatially. They are approximately negatives of each other.

f. Statistics of the relative spatial phase difference $\delta\phi$ between the CDW phase $\phi_i^{\mathrm{C}}(r)$ at $Q_i$ and the PDW phase $\phi_i^{\mathrm{P}}(r)$ at $P_i$ in the coterminous images $g_Q(r, -9\text{ mV})$ and $\delta\Delta_{\mathrm{P}}(r)$. The spatial phase difference, defined as $|\delta\phi_i(r)| \equiv |\phi_i^{\mathrm{P}}(r) - \phi_i^{\mathrm{C}}(r)|$, between all three CDW and PDW at $Q_i:P_i$ is $|\delta\phi|_{\mathrm{RMS}} = 0.96\pi$.

g. Coterminous measurement of CDW $g_Q(r, -9\text{ mV})$ and PDW $\delta\Delta_{\mathrm{P}}(r)$ along an trajectory (arrows in **c,d**).

# Methods for
## Detection of a Pair Density Wave State in UTe$_2$

Qiangqiang Gu, Joseph P. Carroll, Shuqiu Wang, Sheng Ran, Christopher Broyles, Hasan Siddiquee, Nicholas P. Butch, Shanta R. Saha, Johnpierre Paglione, J.C. Séamus Davis and Xiaolong Liu

**Materials and Methods**

**CDW visualization in non-superconductive UTe$_2$**

*Differential conductance imaging of CDW at T = 4.2 K*
At $T$ = 4.2 K and using superconducting-tips to study the UTe$_2$ (0-11) surface, we measure differential tunneling conductance spectra $g(\boldsymbol{r}, V)$ to visualize the CDW in the normal state of UTe$_2$. Extended data Figs. 1a-d show $g(\boldsymbol{r}, V)$ images $V$ = -7 mV, -15 mV, -23 mV and -29 mV with Fourier transform $g(\boldsymbol{q}, V)$ shown as Extended Data Figs. 1e-h. Three CDW peaks at $\boldsymbol{Q}_{1,2,3}$ occur in all $g(\boldsymbol{q}, V)$ representing incommensurate charge density modulations with energy scale up to at least ~30 meV consistent with Ref. 24.

*CDW Visualization at Incommensurate Wavevectors $\boldsymbol{Q}_{1,2,3}$*
To calculate the amplitude $g_{\boldsymbol{Q}_i}(\boldsymbol{r})$ of the CDW modulation represented by the peaks at $\boldsymbol{Q}_i$ ($i = 1,2,3$), we apply a two-dimensional computational lock-in technique. Here $g(\boldsymbol{r})$ is multiplied by the term $e^{i\boldsymbol{Q}_i \cdot \boldsymbol{r}}$, and integrated over a Gaussian filter to obtain the lock-in signal

$$g_{\boldsymbol{Q}_i}(\boldsymbol{r}) = \frac{1}{\sqrt{2\pi}\sigma_r} \int d\boldsymbol{R}\, g(\boldsymbol{R}) e^{i\boldsymbol{Q}_i \cdot \boldsymbol{R}} e^{\frac{|\boldsymbol{r}-\boldsymbol{R}|^2}{2\sigma_r^2}} \tag{M1}$$

where $\sigma_r$ is the cut-off length in the real space. In $\boldsymbol{q}$-space this lock-in signal is

$$g_{\boldsymbol{Q}_i}(\boldsymbol{r}) = \mathcal{F}^{-1} g_{\boldsymbol{Q}_i}(\boldsymbol{q}) = \mathcal{F}^{-1}[\mathcal{F}(g(\boldsymbol{r}) e^{i\boldsymbol{Q}_i \cdot \boldsymbol{r}}) \cdot \frac{1}{\sqrt{2\pi}\sigma_Q} e^{-\frac{q^2}{2\sigma_Q^2}}] \tag{M2}$$

where $\sigma_Q$ is the cut-off length in $\boldsymbol{q}$-space. Here $\sigma_r = 1/\sigma_Q$. Next, $g_Q(\boldsymbol{r}, V)$, the inverse Fourier transform of the combined $\boldsymbol{Q}_i$ ($i = 1,2,3$) CDWs, is presented in Extended Data Figs. 1i-l.

To specify the effect of filter size on the inverse Fourier transform we show in Extended Data Figs. 1m-t the real space density of states $g(\boldsymbol{r}, 10\,\text{mV})$, its Fourier transform $g(\boldsymbol{q}, 10\,\text{mV})$, and the evolution of inverse Fourier transform images as a function of the real-space cut-off length $\sigma_r$. The differential conductance map $g_Q(\boldsymbol{r}, 10\,\text{mV})$ is displayed at a series of $\sigma_r$ including 10 Å, 12 Å, 14 Å, 18 Å, 24 Å and 35 Å. The distributions of the CDW domains in the filtered $g_Q(r, 10 \text{ mV})$ images with a cut-off length of 10 Å, 12 Å, 14 Å, 18 Å, 24 Å are highly similar. The cut-off length used in Main Text Fig. 2d is 14 Å such that the domains of the CDW modulations are resolved, the irrelevant image distortions are excluded. The same filter size of 14 Å is chosen for all three $Q_i$ vectors. Formally the equivalent inverse Fourier transform analysis is carried out for main text Figs. 4c,d, but with a filter size of 11.4 Å to filter both the CDW and PDW peaks.

*Simulated UTe₂ topography*

To identify $q$-space peaks resulting from the (0-11) cleave plane structure of UTe₂ we simulate the topography of the UTe₂ cleave plane and Fourier transform this simulation. Subsequently we can distinguish clearly the CDW signal from the structural periodicity of the surface. The simulation is calculated based on the ideal lattice constant of the (0-11) plane of the UTe₂, $a^* = 4.16$ Å and inter-Te-chain distance $b^* = 7.62$ Å. Extended Data Fig. 2a is a simulated $T(r)$ image in FOV of 14.5 nm. The simulated topography $T(r)$ is in a good agreement with experimental $T(r)$ images presented throughout. The Fourier transform, $T(q)$, of the simulated $T(r)$ in Extended Data Fig. 2b shows six sharp peaks, confirming that they are the primary peaks resulting from the cleave plane structure. Most importantly, the CDW peaks in Main Text Fig. 2c are not seen in the simulation. They are therefore not caused by the surface periodicity but instead originate from the electronic structure, as first demonstrated in Ref. 24.

**Normal-tip PDW detection at the NIS gap-edge of UTe₂**

Initial STM searches for a PDW on UTe₂ were carried out using a normal-tip at 280 mK. Extended Data Fig. 3a displays a typical line cut of $dI/dV|_{\text{NIS}}$ spectrum taken from the FOV in Extended Data Fig. 3b. There is a large residual density of states (DOS) near the Fermi level. The gap depth $H$ is defined as the difference between the gap bottom in the $dI/dV|_{\text{NIS}}$ spectrum and the coherence peak height, i.e., $H \equiv dI/dV|_{\text{NIS}}(V \equiv \Delta_{\text{UTe}_2}) - dI/dV|_{\text{NIS}}(V \equiv 0)$. Its modulation is extracted from the $dI/dV|_{\text{NIS}}$ linecut and presented in Extended Data Fig. 3c; it modulates perpendicular to the Te atom chains.

Conventional, normal-insulator-superconductor (NIS) tunneling does reveal superconducting energy-gap modulations as shown in Extended Data Fig. 3a. The superconducting energy gap is defined as half of the peak-to-peak distance in the $dI/dV|_{\text{NIS}}$ spectrum (Fig. 3a and Extended Data Fig. 3d). Its magnitude $|\Delta_{\text{UTe}_2}|$ is found to lie approximately between 250 µeV and 300 µeV. We measure variations in the coherence peak position from the $dI/dV|_{\text{NIS}}$ spectrum at each location $r$. The two energy maxima near $\Delta_{\text{UTe}_2}$ of each $dI/dV|_{\text{NIS}}$ spectrum are fitted with a second-order polynomial function ($R^2_{RMS} =$ 0.87). The energy gap is defined as the maxima of the fit, $\Delta_+$ for $V > 0$ and $\Delta_-$ for $V < 0$. The total gap map $\Delta_{\text{UTe}_2}(\boldsymbol{r}) \equiv [\Delta_+(\boldsymbol{r}) - \Delta_-(\boldsymbol{r})]/2$ is derived from $\Delta_+$ and $\Delta_-$ (Extended Data Fig. 3e). The Fourier transform of $\Delta_{\text{UTe}_2}(\boldsymbol{r})$, $\Delta_{\text{UTe}_2}(\boldsymbol{q})$ (Extended Data Fig. 3f) reveals three peaks at wavevectors $\boldsymbol{P}_{i=1,2,3}$. They are the initial signatures of the energy gap modulations of the three coexisting PDW states in UTe2.

**Superconductive-tip PDW visualization at the SIS gap-edge of UTe2**

*Tip preparation*
Atomic-resolution Nb superconducting-tips are prepared by field emission. To determine the tip gap value during our experiments, we measure conductance spectrum on UTe2 at 1.5 K ($T_c$ = 1.65 K), where the UTe2 superconducting gap is closed. The tip gap $|\Delta_{\text{tip}}| \cong 1.37$ meV is the energy of the coherence peak (Extended Data Fig. 4a). The measured spectrum is fitted using a Dynes model[40]. The typical $dI/dV|_{\text{SIS}}$ measured at 280 mK on UTe2 (Main Text Fig. 3b) shows the total gap value $E = |\Delta_{\text{tip}}| + |\Delta_{\text{UTe}_2}| \approx 1.62$ meV. Therefore, we estimate $|\Delta_{\text{UTe}_2}| \approx 0.25$ meV, consistent with the previous reports (Ref. 9), the $dI/dV|_{\text{NIS}}$ shown in Main Text Fig. 3a and Extended Data Fig. 3.

*SIS tunneling amplification of energy gap measurements*
To determine the energy of $E_+(\boldsymbol{r})$ and $E_-(\boldsymbol{r})$ at which the maximum conductance in $dI/dV|_{\text{SIS}}(V)$ occurs, we fit the peak of the measured $dI/dV|_{\text{SIS}}(V)$ spectra using a second-order polynomial fit (Equation M3).
$$g(V) = aV^2 + bV + c \qquad (M3)$$
This polynomial fits excellently with the experimental data. Extended Data Figs. 4b,c show two typical $dI/dV|_{\text{SIS}}(V)$ spectra measured at $+V$ and $-V$ along the trajectory indicated in Fig. 3c. The evolution of fits $g(V)$ in Extended Data Figs. 4d,e show a very clear energy gap modulation.

*Shear correction and Lawler-Fujita algorithm*
To register multiple images to precisely the same field of view, a shear correction technique and the Lawler-Fujita (LF) algorithm are implemented to the experimental data. Then, to recover the arbitrary hexagon of the Te lattice, shear correction is applied to correct any image distortions caused by long-range scanning drift during days of continuous measurement.

To correct against picometer-scale distortions of the lattice we apply the LF algorithm. Let $\tilde{T}(\tilde{r})$ represent a topograph of a perfect UTe$_2$ lattice without any distortion. Three pairs of Bragg peaks $\boldsymbol{Q}_1$, $\boldsymbol{Q}_2$ and $\boldsymbol{Q}_3$ can be obtained from FT of $\tilde{T}(\tilde{r})$. Hence $\tilde{T}(\tilde{r})$ is expected to take the form

$$\tilde{T}(\tilde{r}) = \sum_{i=1}^{3} T_i \cos(\boldsymbol{Q}_i \cdot \tilde{r} + \theta_i), \tag{M4}$$

The experimentally obtained topography $T(r)$ may suffer from a slowly varying position-dependent spatial phase shift $\theta_i(r)$, which can be given by

$$T(r) = \sum_{i=1}^{3} T_i \cos(\boldsymbol{Q}_i \cdot r + \theta_i(r)), \tag{M5}$$

To get $\theta_i(r)$, we employ a computational two-dimensional lock-in technique to the topography

$$A_{\boldsymbol{Q}}(r) = \int d\boldsymbol{R}\, T(\boldsymbol{R}) e^{i\boldsymbol{Q}\cdot\boldsymbol{R}} e^{-\frac{(r-R)^2}{2\sigma^2}}, \tag{M6}$$

$$A_{\boldsymbol{Q}_i}(r) = \mathcal{F}^{-1} A_{\boldsymbol{Q}_i}(q) = \mathcal{F}^{-1}[\mathcal{F}(T(r) e^{i\boldsymbol{Q}_i \cdot r}) \cdot \frac{1}{\sqrt{2\pi}\sigma_Q} e^{-\frac{q^2}{2\sigma_Q^2}}], \tag{M7}$$

$$|A_{\boldsymbol{Q}}(r)| = \sqrt{\left(Re A_{\boldsymbol{Q}}(r)\right)^2 + \left(Im A_{\boldsymbol{Q}}(r)\right)^2}, \tag{M8}$$

$$\theta_i(r) = \tan^{-1} \frac{Im A_{\boldsymbol{Q}}(r)}{Re A_{\boldsymbol{Q}}(r)}. \tag{M9}$$

where $\sigma$ is chosen to capture the lattice distortions. In the Lawler-Fujita analysis, we use $\sigma_q = 3.8\, nm^{-1}$. Mathematically, the relationship between the distorted and the perfect lattice for each $\boldsymbol{Q}_i$ is $\boldsymbol{Q}_i \cdot r + \theta_i(r) = \boldsymbol{Q}_i \cdot \tilde{r} + \theta_i$. We define another global position-dependent quantity, the displacement field $\boldsymbol{u}(r) = r - \tilde{r}$, which can be obtained by solving equations

$$\boldsymbol{u}(r) = \begin{pmatrix} \boldsymbol{Q}_1 \\ \boldsymbol{Q}_2 \\ \boldsymbol{Q}_3 \end{pmatrix}^{-1} \begin{pmatrix} \theta_1 - \theta_1(r) \\ \theta_2 - \theta_2(r) \\ \theta_3 - \theta_3(r) \end{pmatrix}. \tag{M10}$$

Finally, a drift-corrected topography, $\tilde{T}(\tilde{r})$ is obtained by

$$\tilde{T}(\tilde{r}) = T(r - \boldsymbol{u}(r)). \tag{M11}$$

By applying the same correction of $\boldsymbol{u}(r)$ to the simultaneously taken differential conductance map $g(r)$, we can get

$$\tilde{g}(\tilde{r}) = g(r - \boldsymbol{u}(r)). \tag{M12}$$

where $\tilde{g}(\tilde{r})$ is the drift-corrected differential conductance map.

*Lattice registration of UTe$_2$ energy gap $\Delta_{UTe_2}(r)$*

We measure two separate $dI/dV|_{SIS}(r, V)$ maps separated by several days, and in two overlapping FOVs, with energy ranges -1.68 meV < $E$ < -1.48 meV and 1.5 meV < $E$ < 1.7 meV.

Therefore, we obtain two datasets, $T_+(\mathbf{r})$ with the simultaneous $dI/dV|_{\mathrm{SIS}+}(\mathbf{r}, V)$ at positive bias and $T_-(\mathbf{r})$ with the simultaneous $dI/dV|_{\mathrm{SIS}-}(\mathbf{r}, V)$ at negative bias.

After the shear and LF correction are applied, the lattice in the corrected topographs of $T_+(\mathbf{r})$ and $T_-(\mathbf{r})$ become nearly perfectly periodic. Next, we perform rigid spatial translations to register the two topographs to the exact same FOV with a lateral precision better than 27 pm. Extended Data Figs. 5a,b show two topographs of registered $T_+(\mathbf{r})$ and $T_-(\mathbf{r})$. Cross-correlation (XCORR) of two images $I_1$ and $I_2$, $X(\mathbf{r}, I_1, I_2)$ at $\mathbf{r}$ is obtained by sliding two images $\mathbf{r}$ apart and calculating the convolution,

$$X(\mathbf{r}, I_1, I_2) = \frac{\int I_1^*(\mathbf{r}_1) I_2(\mathbf{r}+\mathbf{r}_1) d\mathbf{r}_1}{\sqrt{\int |I_1(\mathbf{r}_1)|^2 d\mathbf{r}_1 \int |I_2(\mathbf{r}_2)|^2 d\mathbf{r}_2}}. \tag{M13}$$

where the denominator is a normalization factor such that when $I_1$ and $I_2$ are exactly the same image, we can get $X(\mathbf{r}=\mathbf{0}, I_1, I_2) = 1$ with the maximum centered at (0,0) cross-correlation vector. Extended Data Fig. 5c shows the maximum of XCORR between $T_+(\mathbf{r})$ and $T_-(\mathbf{r})$ coincides with the (0,0) cross-correlation vector. The offset of the two registered images are within one pixel. The multiple-image registration method is better than 0.5 pixel = 27 pm in the whole FOV, and the maxima of the cross-correlation coefficient between the topographs is 0.93. All transformation parameters applied to $T_+(\mathbf{r})$ and $T_-(\mathbf{r})$ to yield the corrected topographs are subsequently applied to the corresponding $dI/dV|_{\mathrm{SIS}}(\mathbf{r}, V)$ maps obtained at positive and negative voltages.

*Particle-hole symmetry of the superconducting energy gap $\Delta_{\mathrm{UTe}_2}(\mathbf{r})$*

The cross-correlation map in Extended Data Fig. 5f provides a two-dimensional measure of agreement between the positive and negative $dI/dV|_{\mathrm{SIS}}(V)$ energy-maxima maps in Extended Data Figs. 5d,e. The inset of Extended Data Fig. 5f shows a linecut along the trajectory indicated in Extended Data Fig. 5f. It shows a maximum of 0.92 and coincides with the (0,0) cross-correlation vector. Thus, it shows that gap values between positive bias and negative bias are highly correlated.

*PDW Visualization at incommensurate wavevectors $P_{1,2,3}$*

The inverse Fourier transform analysis for PDW state in Main Text Fig. 4c is implemented using the same technique described in Methods Section. The filter size chosen to visualize the PDW is 11.4 Å. The inverse Fourier transform of the CDW in Main Text Fig. 4d is calculated using the identical filter size of 11.4 Å.

*Independent PDW visualization experiments*

To confirm that the PDW discovered is present in multiple FOVs we show a typical example of the gap modulation $\Delta_+(r)$ from one different field of view in Extended Data Fig. 6. The $dI/dV|_{SIS}(r,V)$ map is measured in the voltage region surrounding the positive Nb-UTe$_2$ energy maxima near 1.6 meV. The spectra in this FOV are fitted with a second-order polynomial and shear corrected as described in Methods. The resulting gap map, $\delta\Delta_+(r)$ is presented in Extended Data Fig. 6b. The Fourier transform of this map, $\delta\Delta_+(q)$ is presented in Extended Data Fig. 6c. $\delta\Delta_+(q)$ features the same PDW wavevectors, $(P_1, P_2, P_3)$ reported in the main text.

**Energy modulations of subgap Andreev resonances**

Surface Andreev bound states (SABS) must occur in *p*-wave topological superconductors[41]. Moreover, based on the phase changing quasiparticle reflections at the *p*-wave surface, finite-energy Andreev resonances should also occur in the junction between a *p*-wave and an *s*-wave superconductor[42] and are observed in UTe$_2$. Inside the SIS gap, we measure the $dI/dV|_{SIS}(r,V)$ map in the energy range from -500 μeV to 500 μeV. The map is measured in the FOV in Extended Data Fig. 7a, the same FOV as Figs. 3,4. Three conductance peaks are resolved at approximately -300 μeV, 0 and 300 μeV, annotated with green arrows in the typical subgap spectrum in Extended Data Fig. 7b. The energy-maximum of the positive subgap states between 200 μeV to 440 μeV is assigned as $A_+$. The energy-maximum of the negative subgap states between -440 μeV to -200 μeV is assigned as $A_-$. The averaged energy of the Andreev subgap states is defined as $\Delta_A(r) \equiv [A_+(r)-A_-(r)]/2$, which ranges from 300 μeV to 335 μeV (Extended Data Fig. 7c). Fourier transform of the subgap energies $\Delta_A(q)$ exhibit two sharp peaks at the PDW wavevectors $P_1$ and $P_2$ (Extended Data Fig. 7d).

In the case of two superconductors with very different gap magnitude, when the sample bias voltage shifts the smaller gap edge to the chemical potential, the Andreev process of electron (hole) transmission and hole (electron) reflectional plus electron-pair propagation may produce an energy-maximum in d$I$/d$V$|$_{SIS}$ at the voltage of smaller gap energy. Hence the observations in Extended Data Fig. 7d may be expected if the UTe$_2$ superconducting energy gap is modulating at the wavevectors $P_1$ and $P_2$. Extended Data Fig. 7e shows the energy of the Andreev states modulate in space with a peak-to-peak amplitude near 10 μeV (see histogram in Extended Data Fig. 7f).

**Enhancement of SNR using superconductive tips**

Superconducting STM tips provide an effective energy resolution beyond the Fermi-Dirac limit. They have therefore been widely used as a method of enhancing the energy resolution of STM spectra[26-31].

To better quantify the SNR improvement of the measured energy gap modulations we compare the fitting quality of the superconducting gap maps obtained using a normal-tip (Extended Data Fig. 3) and a superconducting-tip (Fig. 4). The fitting quality is defined using the coefficient

$$R^2(\boldsymbol{r}) = 1 - \frac{\sum_{i=1}^{N}[g(\boldsymbol{r},V_i) - dI/dV(\boldsymbol{r},V_i)]^2}{\sum_{i=1}^{N}[g(\boldsymbol{r},V_i) - \bar{g}(\boldsymbol{r})]^2} \quad (M14)$$

where $dI/dV(V)$ is the measured spectrum, $g(\boldsymbol{r}, V)$ is the fitted spectrum and $\bar{g}(\boldsymbol{r})$ is the averaged fitted spectrum. Extended Data Fig. 8a shows a typical spectrum measured using a superconductive tip, $dI/dV|_{SIS}$ from the FOV in Main Text Fig. 3c. Extended Data Fig. 8d is a typical $dI/dV|_{NIS}$ spectrum measured using a normal-tip from the FOV in Extended Data Fig. 3. The energy-maximum noise level is decisively lower in $dI/dV|_{SIS}$ spectra than in $dI/dV|_{NIS}$ spectra and the fitting quality $R^2_{SIS}$ is significantly higher than $R^2_{NIS}$.

Extended Data Figs. 8b,c are maps of the fitting parameter $R^2$ calculated from fitting the $dI/dV|_{SIS}$ energy maxima map obtained using a superconductive tip, i.e., the $\Delta_{UTe_2}(\boldsymbol{r})$ images presented in Main Text Figs. 3e,f. Extended Data Figs. 8e,f are maps of the $R^2$ calculated from the coherence peak fitting of $dI/dV|_{NIS}$ obtained using a normal tip, i.e., the $\Delta_{UTe_2}(\boldsymbol{r})$ images presented in Extended Data Fig. 3e. Comparing these $R^2$ quality-of-fit parameter maps one finds that a much larger fraction of normal-tip coherence peak maps have poor correspondence with the fitting procedures used. For superconducting tips the root-mean square (RMS) values of the fitting parameter, $R^2_{RMS}$, are 0.98 and 0.99 for the positive and negative coherence peak fitting respectively. The normal-tip $R^2_{RMS}$ values are 0.87 and 0.86 for the positive and negative coherence peak fitting respectively. The superconducting tip therefore demonstrably achieves a significant SNR enhancement for evaluation of $\Delta_{UTe_2}(\boldsymbol{r})$ images.

As the SNR is increased in the SIS-convoluted coherence peaks measured using a superconducting tip, it has been possible to resolve the UTe$_2$ energy gap modulations of order ~ 10 μV. Fundamentally the energy resolution is associated with the superconductive tip's ability to resolve the energy at which the *dI/dV*|SIS coherence peak reaches its maximum amplitude. Consequently we determine our energy resolution to be 10 μV.

Thus, the same superconductor energy-gap modulations in $\Delta_{\text{UTe}_2}(\boldsymbol{r})$ of UTe2 can be observed using either a superconducting tip or a normal tip. However, the former significantly increases the SIS conductance at $|E| = \Delta_{\text{UTe}_2} + \Delta_{\text{tip}}$ and allows for considerably better imaging of these energy-maxima and thus $\Delta_{\text{UTe}_2}(\boldsymbol{r})$.

**Interplay of subgap quasiparticles and PDW**

Here we show simultaneous normal-tip measured modulations of the UTe2 subgap states and $\Delta_{\text{UTe}_2}(\boldsymbol{r})$ at *T* = 280 mK, to study their interplay. Extended Data Fig. 9a shows the integrated differential conductance from −250 μV to 250 μV, $\sum_{-250\,\mu V}^{250\,\mu V} g(\boldsymbol{r}, E)$. Inverse Fourier transform of the three wavevectors $\boldsymbol{Q}_{1,2,3}$ from $\sum_{-250\,\mu V}^{250\,\mu V} g(\boldsymbol{r}, E)$ and $P_{1,2,3}$ from the simultaneous $\Delta_{\text{UTe}_2}(\boldsymbol{r})$ in Extended Data Fig. 3e are compared in Extended Data Figs. 9 c,d. Clearly, from the highly distinct spatial structure of these images, there is no one-to-one correspondence between the subgap density of states modulations and the simultaneously measured PDW energy gap modulations in UTe2. Overall there is a very weak anticorrelation with cross-correlation value of -0.23±0.05 that is not inconsistent with coincidence. Hence we demonstrate that there is no deterministic influence of the subgap density of states modulations on the PDW energy gap modulations in superconducting UTe2.

**Visualizing the interplay of PDW and CDW in UTe2**

The analysis of phase difference between PDW and CDW at three different wavevectors is shown in Extended Data Fig. 10. The inverse Fourier transforms of each CDW and PDW wavevector demonstrate a clear half-period shift between the two density waves (Extended Data Figs. 10a-f). This shift motivates the statistical analysis of the phase difference. The phase map of $g_{Q_1}(\boldsymbol{r}, -9 \text{ mV})$, $\phi_1^C(\boldsymbol{r})$, and the phase map of $\Delta_{P_1}(\boldsymbol{r})$, $\phi_1^P(\boldsymbol{r})$, are calculated. The phase difference between two corresponding maps is defined as $|\delta\phi_1| = \phi_1^C(\boldsymbol{r}) - \phi_1^P(\boldsymbol{r})$ for the $\boldsymbol{P}_1 : \boldsymbol{Q}_1$ wavevectors. The identical procedure is carried out for $\boldsymbol{P}_2 : \boldsymbol{Q}_2$ and $\boldsymbol{P}_3 : \boldsymbol{Q}_3$. The histograms resulting from this procedure show that the statistical distributions of the phase shift $|\delta\phi_i|$ are centered around $\pi$ (Extended Data Figs. 10j-l). Although the distribution varies, this $\pi$ phase shift reinforces the observation of the spatial anti-correlation between CDW and PDW.

As shown in inset of Extended Data Fig. 10g the three PDW wavevectors are related by reciprocal lattice vectors: $P_2=P_1-G_3$ and $P_3=G_1-P_1$. Nevertheless the three UTe$_2$ PDWs appear to be independent states when analyzed in terms of the spatial modulations of the amplitude of the $P_{1,2,3}$ peaks from Fig. 4 using equation M8. The amplitude of $P_{1,2}$ has a domain width beyond 10 nm in the real space (Extended Data Figs. 10g,h). The amplitude of $P_3$ is short-ranged of which the averaged domain width is approximately 5 nm (Extended Data Fig. 10i). The one-pixel shift of $P_3$ from the central axis is within the error bar of experimental measurements. The spatial distributions of the three PDWs are negligibly correlated with cross-correlation values of their amplitude of $X(P_1, P_2) = -0.3$, $X(P_1, P_3) = 0.09$, $X(P_2, P_3) = 0.28$. The weak cross-correlation relationships indicate that the three PDWs are independent orders.

**Acknowledgements** The authors are extremely grateful to V. Madhavan for generous and incisive advice and guidance, on how to execute this project. We acknowledge and thank D. Agterberg, F. Flicker, E. Fradkin, Eun-Ah Kim, S. Simon, J. van Wezel and K. Zhussupbekov for key discussions and theoretical guidance. Research at the University of Maryland was supported by the Department of Energy Award No. DE-SC-0019154 (sample characterization), the Gordon and Betty Moore Foundation's EPiQS Initiative through Grant No. GBMF9071 (materials synthesis), NIST, and the Maryland Quantum Materials Center. Q.G., X.L., J.P.C. and J.C.S.D. acknowledge support from the Moore Foundation's EPiQS Initiative through Grant GBMF9457. J.C.S.D. acknowledge support from the Royal Society under Award R64897. J.P.C. and J.C.S.D. acknowledge support from Science Foundation Ireland under Award SFI 17/RP/5445. S.W. and J.C.S.D. acknowledge support from the European Research Council (ERC) under Award DLV-788932.

**Author Contributions** X.L. and J.C.S.D. conceived the project. S.R., C.B., H.S., S.R.S, N.P.B. and J.P. developed, synthesized and characterized materials; Q.G., J.P.C., S.W. and X.L. carried out the experiments; S.W., J.P.C and Q.G. developed and implemented analysis. X.L. and J.C.S.D supervised the project. J.C.S.D. wrote the paper with key contributions from J.P.C., Q.G., X.L. and S.W. The paper reflects contributions and ideas of all authors.

**Data availability** The data shown in the main figures are available from Zenode at https://doi.org/10.5281/zenodo.7662516.

**Code availability** The code is available to qualified researchers from corresponding authors upon reasonable request.

**Competing interests** The authors declare no competing interests.

**Additional information**

**Correspondence and requests for materials** should be addressed to J.C. Seamus Davis, Shuqiu Wang or Xiaolong Liu

**Reprints and permissions information** is available at http://www.nature.com/reprints.

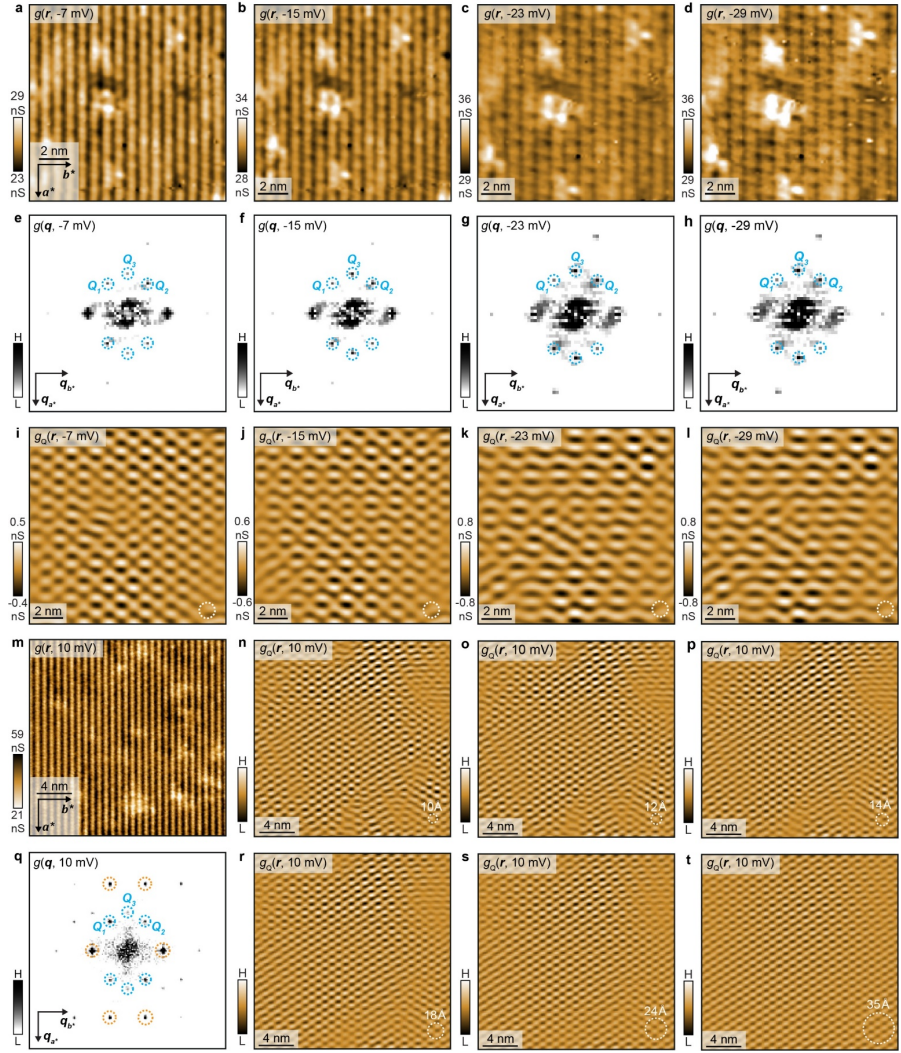

**Extended Data Fig. 1 CDW at different voltages in UTe₂. a-d**. Measured $g(r, V)$ images of UTe$_2$ using normal-STM tips at $T$ = 4.2 K, and at four representative negative sample voltages including -7 mV, -15 mV, -23 mV and -29 mV in the same 12 nm × 12 nm FOV. **e-h**. Fourier transform of the $g(r, V)$ images, $g(q, V)$ at different sample voltages, showing the presence of the three wavevectors corresponding to the CDW order (in dashed blue circles). **i-l.** Inverse Fourier transform the CDW peaks ($Q_1$, $Q_2$, $Q_3$) at different sample voltages. The CDW pattern is independent from the sample voltages for -29 mV < $V$ < -7 mV. A white dashed circle indicates $\sigma$ of Gaussian filter used to isolate CDW peaks in real space. **m-t**. Cut-off dependence of IFT. (**n-p, r-t**) Inverse Fourier transform of CDW peaks $g_Q(r, 10\text{ mV})$ from $g(r, 10\text{ mV})$ in **m** and $g(q, 10\text{ mV})$ in **q**. The images of $g_Q(r, 10\text{ mV})$ are filtered at different cut-off lengths such as 10 Å, 12 Å, 14 Å, 18 Å, 24 Å and 35 Å. The filter size is in the bottom-right corner. $\sigma_r$ chosen for main text Fig. 2d is 14 Å.

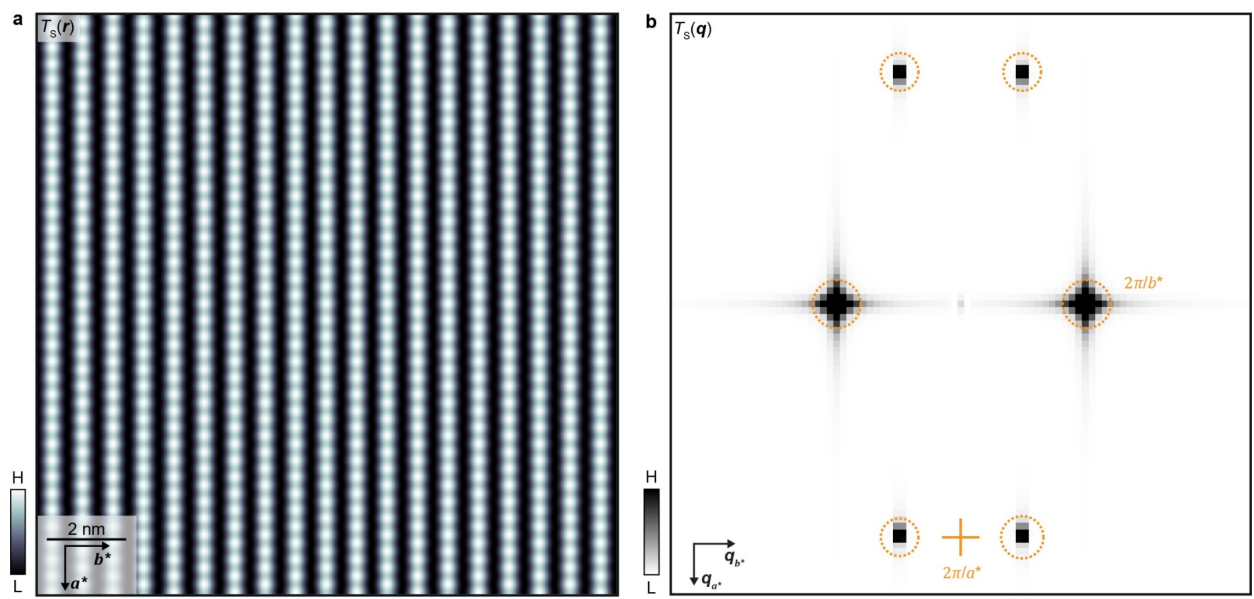

**Extended Data Fig. 2. Simulated topography of UTe$_2$ and its Fourier transform. a.** Simulated topograph, $T_S(\boldsymbol{r})$ of (0-11) cleave surface of UTe$_2$. **b.** Fourier transform of simulated topograph, $T_S(\boldsymbol{q})$. The six primary peaks occur at the reciprocal-lattice wavevectors and are observed in the experimental STM data.

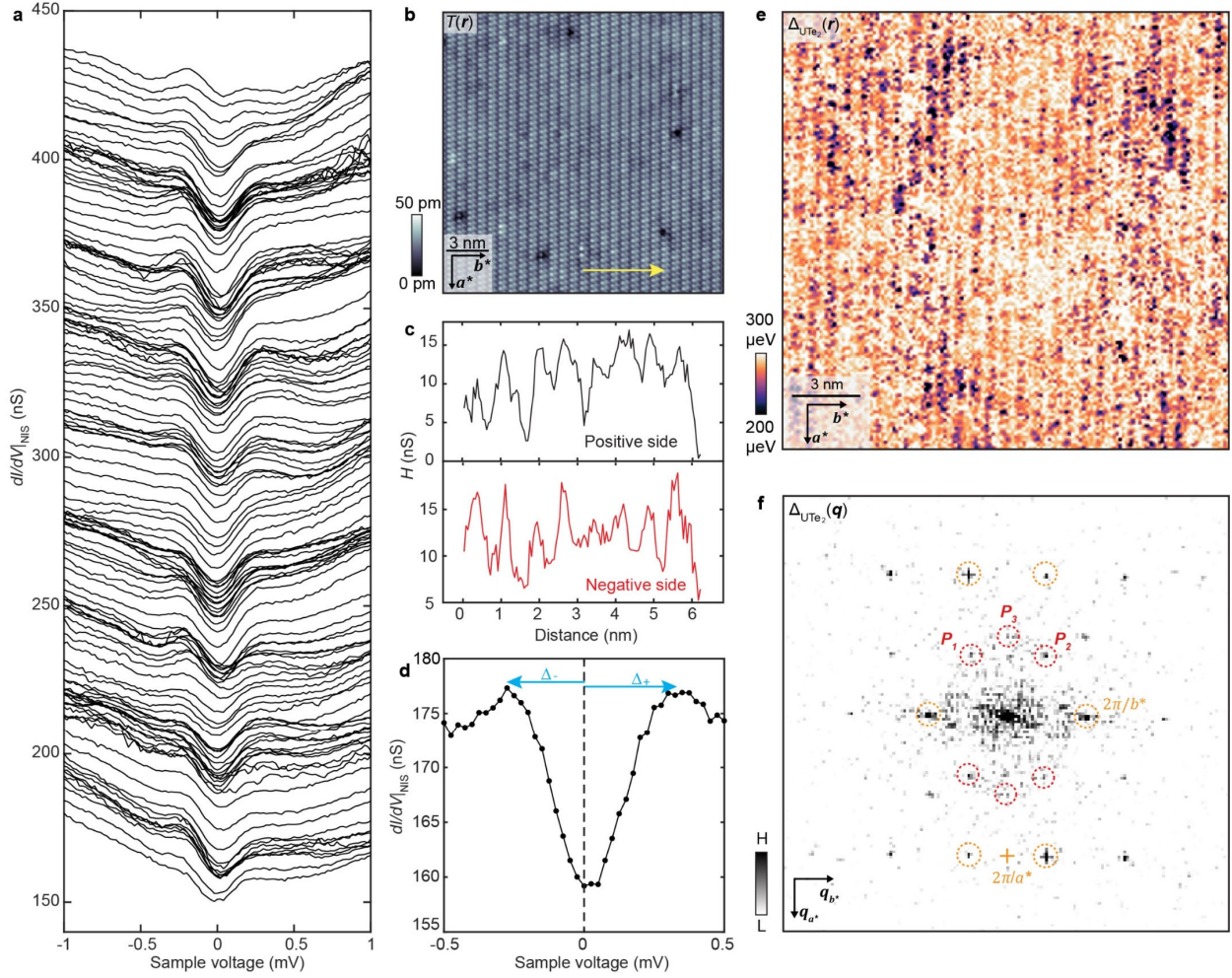

**Extended Data Fig. 3. PDW detection using a normal-tip. a.** A typical line cut of $dI/dV|_{NIS}$ spectra obtained at 280 mK along the trajectory shown in **b** ($I_s$ = 1 nA, $V_s$= -5 mV). **b.** Topograph $T(r)$ obtained using a normal tip. **c.** Gap depth $H$ distribution along the trajectory in **b**. **d.** $dI/dV|_{NIS}$ spectrum displaying the superconducting gap $\Delta^+$ and $\Delta^-$. **e.** Image of half the energy difference between superconducting coherence peaks, i.e. the superconducting energy gap of $\Delta_{UTe_2}(r)$, obtained in the same FOV as **b**, using conventional normal-tip imaging. **f.** $\Delta_{UTe_2}(q)$ the Fourier transform of the $\Delta_{UTe_2}(r)$. Three peaks are seen at the same wavevector as the normal state CDW and indicate the existence of three superconducting PDW states ($I_s$ = 1 nA, $V_s$= -5 mV).

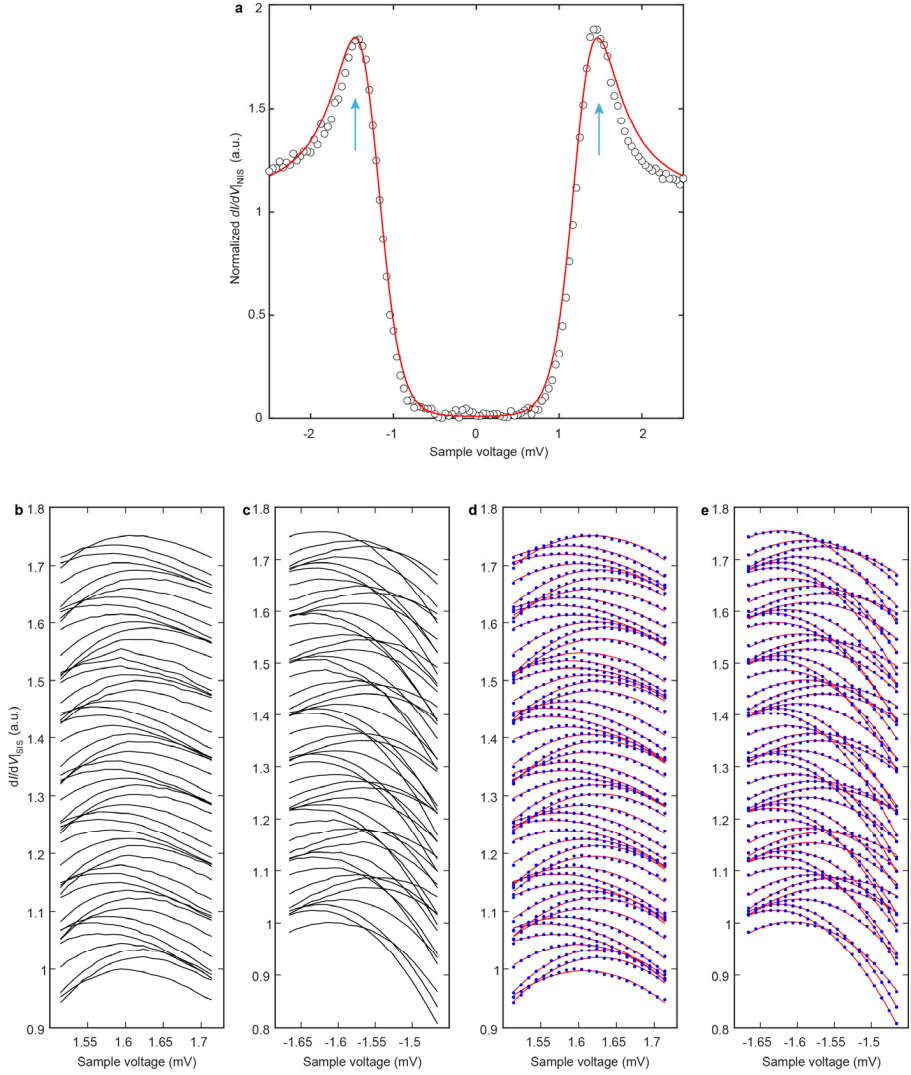

**Extended Data Fig. 4. Determination of the tip gap $\Delta_{\text{tip}}$ and evolution of $dI/dV|_{\text{SIS}}$ spectra with parabolic fitting. a.** A typical spectrum measured on UTe₂ by using a superconducting Nb tip at 1.5 K ($I_s$ = 100 pA, $V_s$ = 4 mV). At this temperature UTe₂ gap is closed, thus the coherence peak value shows the pure Nb tip gap of 1.37 meV. The spectrum is clearly well fitted using the Dynes model. The fitting parameters of the Dynes model are $\Gamma$ = 0.01 meV, $\Delta$ = 1.37 meV. **b, c**. Line cuts of $dI/dV|_{\text{SIS}}(V)$ spectra measured at both negative bias and positive bias along the trajectory shown in main text Fig. 3c. **d, e**. The evolution of $dI/dV|_{\text{SIS}}(V)$ spectra (blue points) from the same data shown in **b, c** and their parabolic fits $g(V)$.

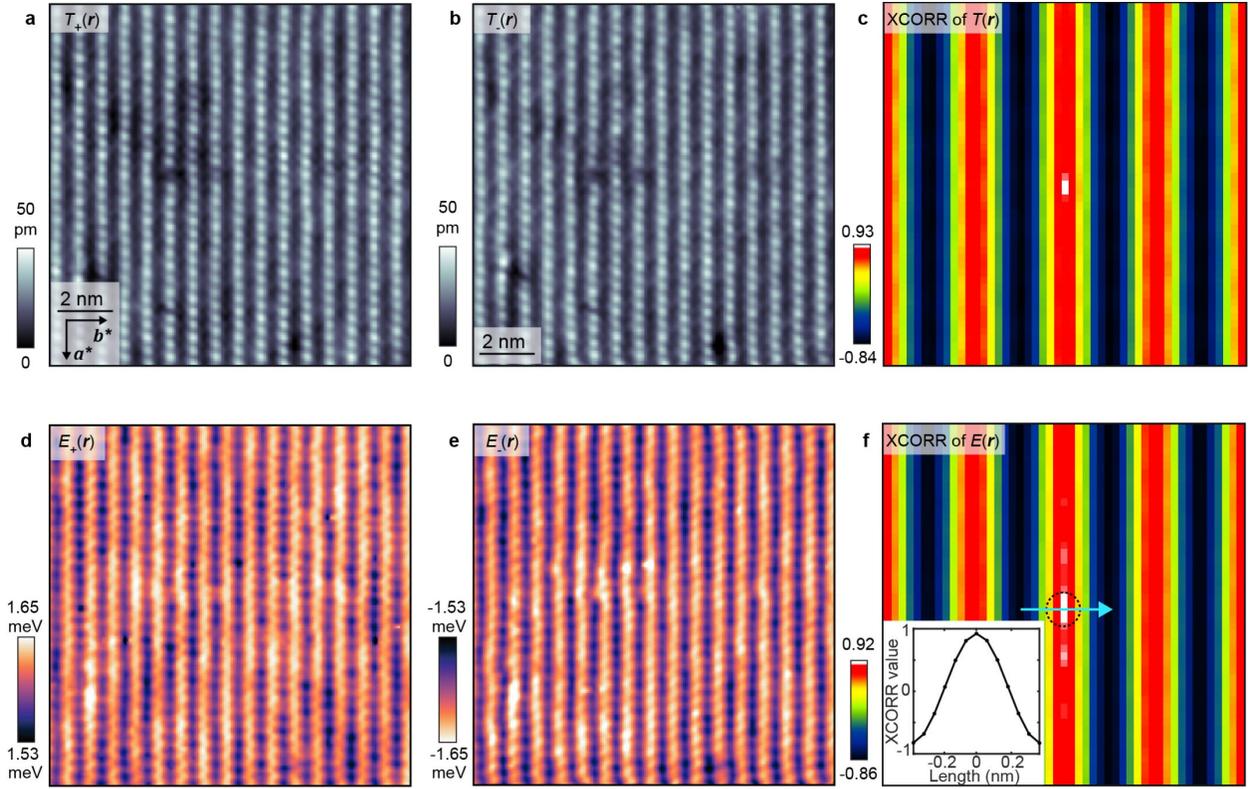

**Extended Data Fig. 5. Spatial registration of topographs and gap maps**. **a, b.** 12 nm × 12 nm topographs after registration. These topographs were obtained concomitantly as $dI/dV|_{SIS}(r,V)$ maps recording positive and negative coherence peaks, respectively. **c.** Cross-correlation (XCORR) map of the registered topographs. The correlation coefficient is 0.93 indicating the two topographs are almost identical. The maxima of the XCORR map is single pixel wide, which suggests a registration precision of 0.5 pixels equivalent to registration precision of 27 pm. **d.** Positive coherence peak map $E_+(r)$ from **a**. **e**. Negative coherence peak map $E_-(r)$ from **b**. **f.** XCORR map providing a two-dimensional measure of correlation between the positive gap map $E_+(r)$ and negative gap map $E_-(r)$. Inset: A linecut along the trajectory indicated in **f**. It shows the maximum is 0.92 and coincides with the (0,0) cross-correlation vector. The strong correlation demonstrates the particle-hole symmetry in superconductive UTe$_2$.

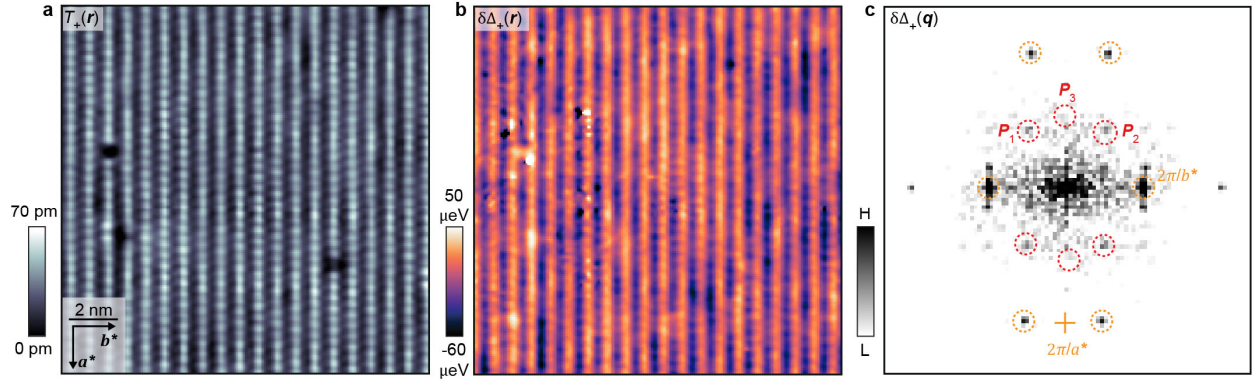

**Extended Data Fig. 6. PDW repeatability analysis. a.** A topograph recorded in a new FOV away from that seen in Main Text Fig. 3c. The image size is 15 nm × 15 nm ($V_s = 3$ mV, $I_s = 2.5$ nA). **b.** $\delta\Delta_+(r)$ map prepared using the same procedure outlined in Methods revealing the same gap modulations as Main Text Fig. 4a. **c.** The Fourier transform of $\delta\Delta_+(r)$ map, $\delta\Delta_+(q)$. ($P_1, P_2, P_3$) PDW peaks are highlighted with dashed red circles and reciprocal lattice vectors highlighted with dashed orange circles.

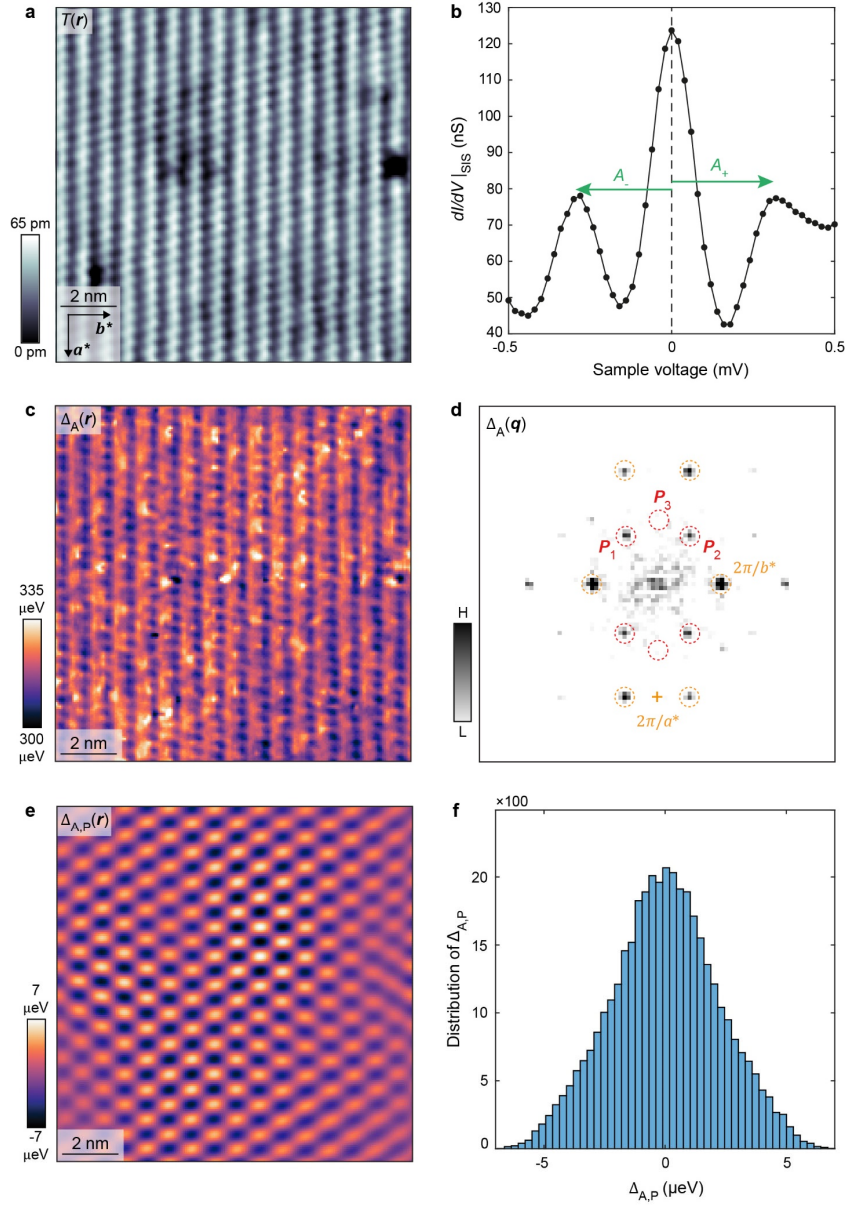

**Extended Data Fig. 7. Imaging of subgap Andreev resonances a.** Topography of the subgap states imaging in the same FOV as Figs. 3,4 in the main text. **b.** A representative $dI/dV|_{SIS}(V)$ spectrum of the subgap states annotated by the green arrows. **c.** Map of the energy scale of the subgap states modulations $\Delta_A(r)$. **d.** Fourier transform of the subgap states modulations $\Delta_A(q)$. $P_{1,2,3}$ PDW peaks are highlighted with dashed red circles **e.** Inverse Fourier transform $\Delta_{A,P}(r)$ of PDW peaks $P_{1,2,3}$. **f.** Histogram of $\Delta_{A,P}(r)$ shows the PDW modulates within 10 μeV.

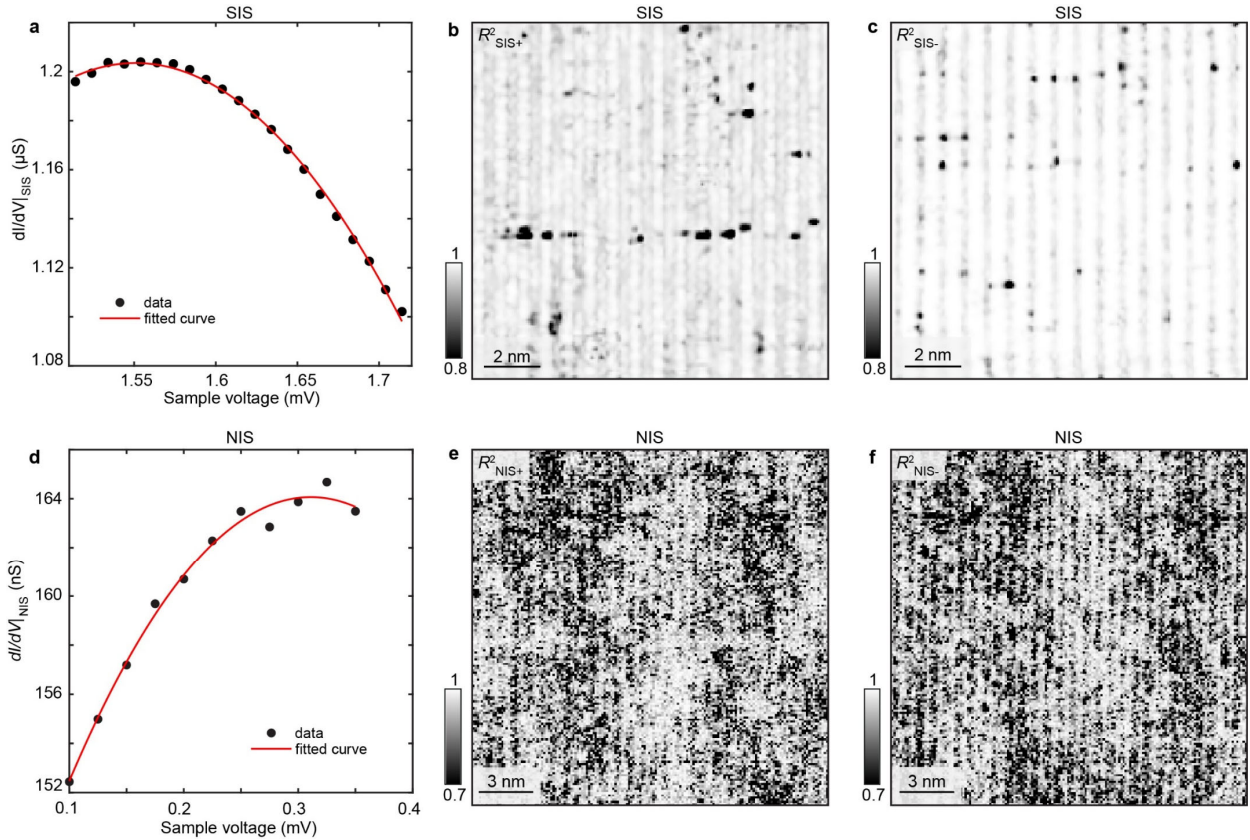

**Extended data Fig. 8. Estimation of SNR using fitting quality of spectra measured with superconductive tips and normal tips**. **a**. Parabolic fit of a typical $dI/dV|_{SIS}$ spectrum measured using superconductive tips. **b, c**. Measured $R^2$ maps used to estimate of the fitting quality of $dI/dV|_{SIS}$ spectra for **b** positive energy and **c** negative energy. The $R^2$ image is from the FOV of Fig. 3c in the main text. **d**. Parabolic fit of a typical $dI/dV|_{NIS}$ spectrum taken using normal tips. **e, f**. Measured $R^2$ maps used to estimate of the fitting quality of $dI/dV|_{NIS}$ spectra for **e** positive energy and **f** negative energy. The $R^2$ image is from the FOV in Extended Data Fig. 3b.

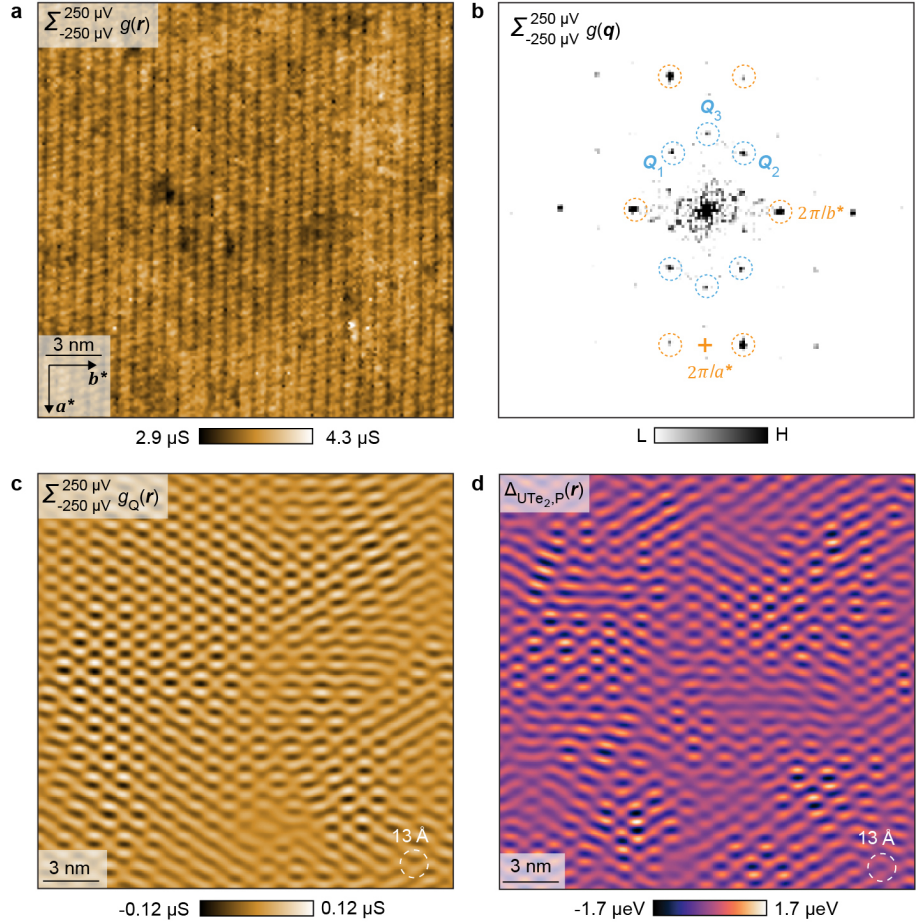

**Extended data Fig. 9. Modulations of subgap states measured using normal tips.** **a**. Sum of all subgap states $\sum_{-250\ \mu V}^{250\ \mu V} g(r, E)$, measured at $T$ = 280 mK. **b**. Fourier transform of subgap states $\sum_{-250\ \mu V}^{250\ \mu V} g(q, E)$ where all three wavevectors $P_{1,2,3}$ are present. **c**. Inverse Fourier transform of $P_{1,2,3}$ from **a**. **d**. Inversed Fourier transform of $P_{1,2,3}$ from Extended Data Figs. 3e,f. The filter size is indicated as a dashed white circle.

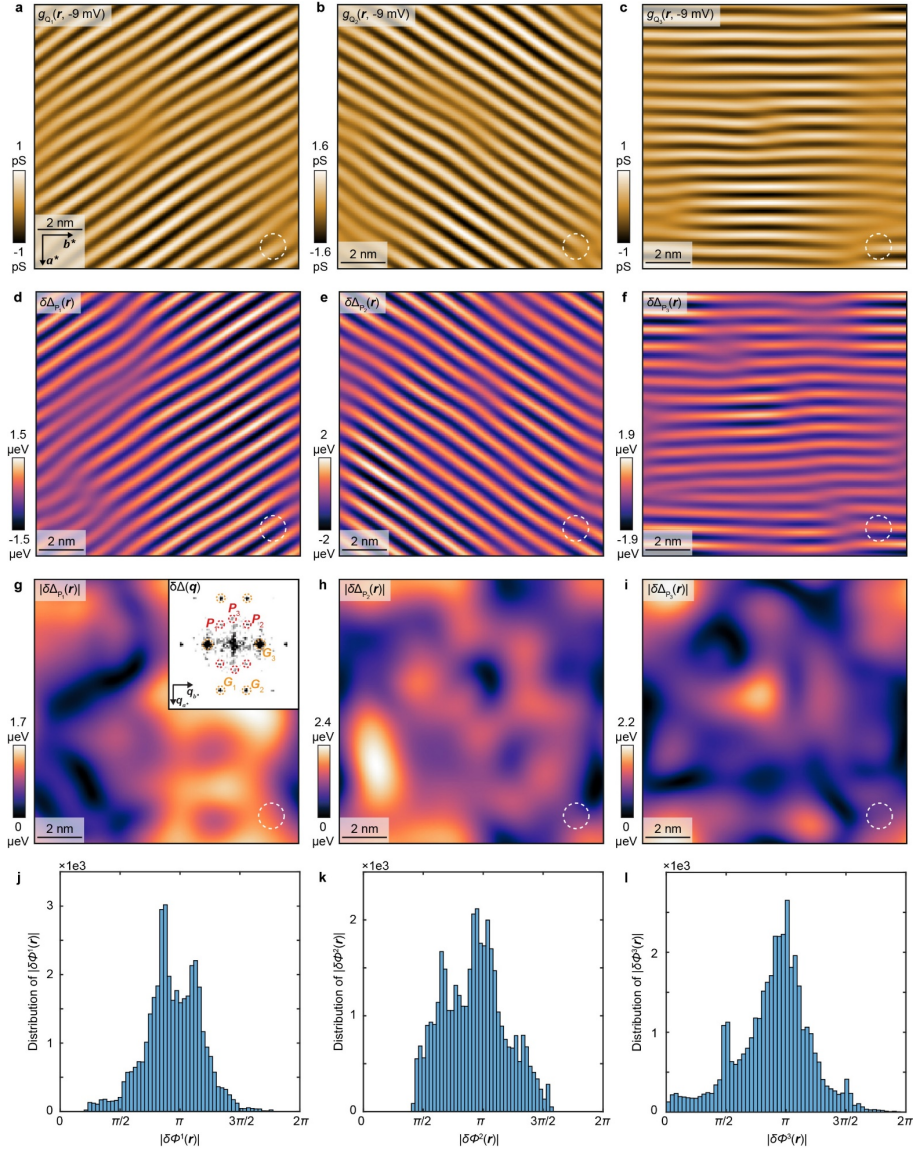

**Extended Data Fig. 10. Phase shift between CDW and PDW. a-c.** Inverse Fourier transforms of the three CDW wavevectors identified $g_{Q_{i=1,2,3}}(r, -9 \text{ mV})$ in the same 12 nm × 12 nm FOV as Main Text Fig. 3c. **d-f.** Inverse Fourier transforms of the three PDW $\delta\Delta_{P_{i=1,2,3}}$ wavevectors in the same FOV as Main Text Fig. 3c. **g-i.** Amplitude for all three PDW wavevectors $P_{i=1,2,3}$. Inset of **g** is the Fourier transform of the energy gap map in which the reciprocal lattice points $G_{i=1,2,3}$ are labelled. **j-l.** Distributions of the relative spatial phase difference $\delta\phi_i(r)$ between $\phi_i^C(r)$ and $\phi_i^P(r)$ from three individual wavevectors. Each histogram is centered around $\pi$ reinforcing the observation of a general phase difference $|\delta\phi_i| \cong \pi$ between the CDW and PDW.